\documentclass[a4paper,unpublished]{quantumarticle}
\pdfoutput=1
\usepackage[utf8]{inputenc}
\usepackage[english]{babel}
\usepackage[T1]{fontenc}
\usepackage{amsmath}
\usepackage{hyperref}
\usepackage{tikz}
\usepackage{lipsum}

\usepackage{graphicx}
\usepackage[justification=justified]{caption}
\usepackage{dcolumn}
\usepackage{bm}

\usepackage{adjustbox}
\usepackage{enumitem}
\usepackage{sty/zongbo}
\usepackage{amsthm}

\newtheorem{theorem}{Theorem}

\newtheorem{proposition}[theorem]{Proposition}

\def\inf{\mathrm{Inf}}
\def\Inf{\inf}

\def\Zfn{\ensuremath{\ZZ_4^n}}

\def\ZfT{\ensuremath{\ZZ_4^T}}
\def\ZfTc{\ensuremath{\ZZ_4^{T^c}}}

\newcommand{\vect}[1]{\ensuremath{\mathrm{vec}\p{ #1 }}}

\begin{document}

\title{Experimental efficient influence sampling of quantum processes}

\author{Hao Zhan}
\thanks{These authors contributed equally to this work}
\affil{National Laboratory of Solid State Microstructures, Key Laboratory of Intelligent Optical Sensing and Manipulation, College of Engineering and Applied Sciences, Jiangsu Physical Science Research Center, and Collaborative Innovation Center of Advanced Microstructures, Nanjing University, Nanjing, 210093, China}
\orcid{0009-0001-9994-1710}

\author{Zongbo Bao}
\thanks{These authors contributed equally to this work}
\affil{State Key Laboratory for Novel Software Technology, New Cornerstone Science Laboratory, Nanjing University, Nanjing 210093, China}
\affil{QuSoft and CWI, Amsterdam, 1098 XG, The Netherlands}
\orcid{0009-0008-9777-7786}

\author{Zekun Ye}
\affil{State Key Laboratory for Novel Software Technology, New Cornerstone Science Laboratory, Nanjing University, Nanjing 210093, China}
\orcid{0009-0001-7465-0458}

\author{Qianyi Wang}
\affil{National Laboratory of Solid State Microstructures, Key Laboratory of Intelligent Optical Sensing and Manipulation, College of Engineering and Applied Sciences, Jiangsu Physical Science Research Center, and Collaborative Innovation Center of Advanced Microstructures, Nanjing University, Nanjing, 210093, China}

\author{Minghao Mi}
\affil{National Laboratory of Solid State Microstructures, Key Laboratory of Intelligent Optical Sensing and Manipulation, College of Engineering and Applied Sciences, Jiangsu Physical Science Research Center, and Collaborative Innovation Center of Advanced Microstructures, Nanjing University, Nanjing, 210093, China}

\author{Penghui Yao}
\email{pyao@nju.edu.cn}
\affil{State Key Laboratory for Novel Software Technology, New Cornerstone Science Laboratory, Nanjing University, Nanjing 210093, China}
\affil{Hefei National Laboratory, Hefei 230088, China.}
\orcid{0000-0002-4104-2069}

\author{Lijian Zhang}
\email{lijian.zhang@nju.edu.cn}
\affil{National Laboratory of Solid State Microstructures, Key Laboratory of Intelligent Optical Sensing and Manipulation, College of Engineering and Applied Sciences, Jiangsu Physical Science Research Center, and Collaborative Innovation Center of Advanced Microstructures, Nanjing University, Nanjing, 210093, China}
\orcid{0000-0002-3247-9189}

\maketitle

\begin{abstract}
Characterizing quantum processes is essential for unlocking the potential of quantum devices. However, standard quantum process tomography is resource-intensive and becomes infeasible on large-scale systems.
Despite alternative approaches have been successfully developed for specific scenarios, they typically rely on
 multi-qubit gates or extensive prior knowledge, limiting their practicability and scalability.
To address these challenges and complement existing approaches, we introduce \emph{influence sampling}, an efficient and scalable protocol that quantifies the \emph{influence} of a quantum process on all qubit subsets using only single-qubit test gates, with sample complexity independent of system size.
Using a photonic platform, we demonstrate influence sampling to identify high-influence qubits, reduce the full process to a smaller effective process, i.e., a junta approximation, and then learn it.
We further confirm scalability by applying the protocol to a 24-qubit system and validate the junta approximation on a two-qubit process.
These results establish influence sampling as a critical characterization technique, facilitating process learning and device assessment.
\end{abstract}

\section{Introduction}
The development of large-scale, multi-qubit quantum computers promises practical quantum advantages beyond classical capabilities~\cite{RN85,doi:10.1126/science.abe8770,RN89,PhysRevLett.134.090601}.
Realizing this potential requires accurate characterization of quantum gates or processes for performance assessment and noise diagnosis across diverse architectures~\cite{RN88,RN94,RN90,RN87,PRXQuantum.6.030202}.
Fully characterizing an $n$-qubit process via quantum process tomography (QPT)~\cite{doi:10.1080/09500349708231894,PhysRevLett.78.390,PhysRevLett.93.080502,RN83} demands $\mathcal{O}(4^n)$ queries, which is prohibitive at scale due to exponential resource consumption~\cite{10.1063/1.4867625}.
Alternative approaches, such as randomized benchmarking~\cite{PhysRevA.77.012307,PhysRevLett.106.180504,PhysRevLett.108.260503,PhysRevLett.109.080505}, classical shadow~\cite{PhysRevLett.127.200501,acharya2023,PhysRevResearch.6.013029}, and quantum verification~\cite{PhysRevA.101.042316,PhysRevLett.114.140505,PhysRevLett.128.020502}, alleviate some costs and show advantages in their specific scenarios, but often rely on multi-qubit gates or extensive prior knowledge~\cite{PhysRevA.102.022410}, limiting their scalability and broad applicability.
This motivates the development of scalable, hardware-friendly protocols that complement existing techniques and speed up characterization.
\par
An $n$-qubit quantum process $\Phi$ is often implicitly treated as acting non-trivially and equally on all qubits, so standard characterization scales with the full system size. 
In practice, however, many processes may act strongly only on a small subset of qubits and are close to the identity elsewhere, behavior reminiscent of a “junta”. 
This motivates the exploration of quantum junta processes~\cite{wang2012property,PhysRevA.84.052328,Thomas2023Testing,bao2023testing,bao2025efficient,chen2024tolerant}, extending Boolean juntas~\cite{FISCHER2004753,CHOCKLER2004301,AS07_qtesting,10.1145/1536414.1536437,Andris2016Efficient,pmlr-v134-chen21b,montanaro2010quantum}: $\Phi$ is a $k$-junta if it acts non-trivially only on a subset $T$ of at most $k$ qubits ($k < n$).
Such processes admit the decomposition $\Phi = \Phi_T \otimes \mathcal{I}_{T^c}$, where $\Phi_T$ acts on the high-influence subset $T$ and $\mathcal{I}_{T^c}$ is the identity on the complement $T^c$. 
Identifying the $\Phi$ as a $k$-junta ($|T| = k$) reduces characterization to a $k$-qubit problem~\cite{Thomas2023Testing,bao2023testing,bao2025efficient}, yielding substantial savings when $n$ is large and $k$ is small.
\par
To identify and harness this latent structure in quantum processes, we introduce \textit{influence sampling}, an efficient, scalable protocol to quantify how strongly a process acts on qubit subsets.
The influence of $\Phi$ on a subset $S$, denoted $\text{Inf}_S[\Phi]$, is defined via the deviation of the process from identity on $S$~\cite{bao2023testing}.
Influence sampling estimates valid upper and lower bounds on $\text{Inf}_S[\Phi]$ for all $2^n$ subsets using only two or three single-qubit test gates, with sample complexity independent of $n$.
By identifying a high-influence subset $T$ with its complement $T^c$, one obtains a junta approximation as $\Phi\approx  \Phi_T \otimes \mathcal{I}_{T^c}$, with an error bound produced by $\text{Inf}_{T^c}[\Phi]$.
Experimentally, we apply influence sampling to a four-qubit photonic system (figure~\hyperref[fig1_schematic_exp]{1a} and figure~\hyperref[fig1_schematic_exp]{1b}) to estimate influence bounds, identify high-influence qubits on different processes comprising both unitary and non-unitary subprocesses, and learn the resulting junta processes.
We then demonstrate scalability by running influence sampling on a 24-qubit system and validate the effectiveness of junta approximation on an imperfect two-qubit identity process.
These results establish influence sampling as a powerful tool for characterizing quantum systems, enriching the quantum benchmarking toolbox and paving the way for efficient learning techniques in scalable quantum technologies.

\section{Influence sampling}
\subsection{Preliminary}
Given an $n$-qubit quantum process $\Phi$, its influence on a qubit set $S\subseteq [n]$, where $[n] = \{1, 2, \dots, n\}$, is defined as follows~\cite{bao2023testing}: 
\begin{equation}
    \inf_{S}[\Phi] = \sum_{\exists i\in S, x_i \neq 0}\chi^{\Phi}_{\mathbf{x}\mathbf{x}},
\end{equation}
where $\mathbf{x}=[x_i]_{i=1}^{n}$ with each $x_i\in \{0,1,2,3\}$, and $\chi^{\Phi} = [\chi^{\Phi}_{\mathbf{x}\mathbf{y}}]$ represents the process ($\chi$-)matrix of $\Phi$~\cite{Nielsen_Chuang_2010}. 
Intuitively, the influence aggregates the diagonal $\chi$-terms corresponding to nontrivial action on $S$. 
For $\Phi$ and $S$, define the reduced subprocess on $S$ by $\Phi_S(\rho_S) = \text{Tr}_{S^c}\left[ \Phi\left(\rho_S \otimes I_{S^c}/2^{|S^c|}\right) \right]$, where $I_{S^c}/2^{|S^c|}$ is the maximally mixed state on the complement $S^c$. 
The influence on $S$ can then be written as $\inf_{S}[\Phi] = 1-F(\Phi_S,\mathcal{I}_{S})$, where $F$ denotes the process fidelity. 
More details are provided in the appendix~\hyperref[app_a]{A}.
\par
Let $T\subseteq [n]$ be a qubit subset with complement $T^c$. The distance between $\Phi$ and the $T$-junta $\Phi_T\otimes\mathcal{I}_{T^c}$, as the junta‑approximation error, is bounded by the influence of $\Phi$ on $T^c$ as
\begin{equation}
D(\Phi_T\otimes\mathcal{I}_{T^c},\Phi)\leq \sqrt{\inf_{T^c}[\Phi]} + \frac{\inf_{T^c}[\Phi]}{\sqrt{2}},
\label{eq_main_the}
\end{equation}
where $\Phi_T$ is the reduced subprocess on $T$ and the distance is defined as $D(\Psi_1,\Psi_2) = \norm{\chi^{\Psi_1}-\chi^{\Psi_2}}_F/\sqrt{2}$ with $\norm{\cdot}_F$ as the Frobenius norm. 
The formal theorem statement and proof are given in appendix~\hyperref[app_b]{B}.
If $T$ is a high-influence subset such that the influence on its complement $\inf_{T^c}[\Phi]$ is sufficiently small, equation~(\ref{eq_main_the}) yields a non‑trivial bound.
\par
\subsection{Influence sampling protocol}
\begin{figure*}[htbp]
    \centering
    \includegraphics[width=1\textwidth]{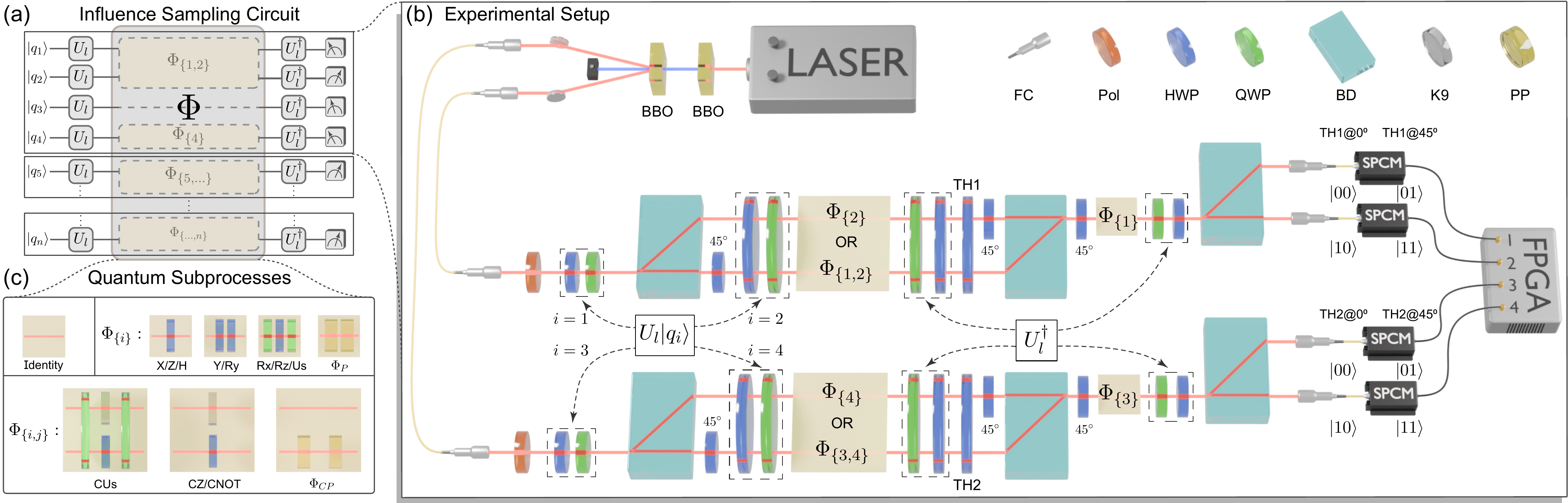}
    \caption{(a) Quantum circuit for influence sampling. The initial qubits are randomly prepared on the computational bases, and the same single-qubit test gate $U_l$, is applied to each qubit, chosen from $\{I,H,R_x\left(\frac{\pi}{2}\right)\}$. Sampling results are obtained by recording which qubits flip relative to their initial state. (b) Experimental setup. Four qubits are encoded on polarization and path DoF of a pair of photons. State preparation and implementation of $U_l$ are merged in a group of wave plates. The target quantum processes $\Phi$ are constructed from some subprocesses $\Phi_{\{i\}}$ and $\Phi_{\{i,j\}}$, as shown in (c). Measurements in the computational basis are performed through four switchable measurement groups controlled by two toggling HWPs (TH1 and TH2). FC: fiber coupler; Pol: polarizer; BD: beam displacer; K9: K9 glass; PP: phase plate;}
    \label{fig1_schematic_exp}
\end{figure*}
We now introduce the protocol to extract the influence of $\Phi$ on any qubit subset $S$, consisting of two stages: quantum sampling and classical post-processing. 
\par
In the quantum sampling stage with the circuit shown in figure~\hyperref[fig1_schematic_exp]{1(a)}, we firstly choose a fixed single-qubit test gate $U_l$ with $l\in\{1,2,3\}$, and $U_1 = I$, $U_2 = H$ and $U_3 = R_x(\frac{\pi}{2})$,
which implement Pauli $Z$-, $X$-, $Y$-basis tests, respectively. 
In each shot, we prepare a random computational basis state $\ket{a}$, with $a$ uniformly sampled from $\{0, 1\}^n$.
After applying the test gate $U_l^{\otimes n}$, going through $\Phi$ and applying $\left(U_l^{\dagger}\right)^{\otimes n}$, computational basis measurements are performed to obtain an $n$-bit measurement outcome $b\in \{0, 1\}^n$. 
We then only record the ``flipped'' position, producing a sampled qubit subset 
\begin{equation}
    \mathcal{T}_l = \{i \in [n] \mid a_i \neq b_i\}.
\end{equation}
Repeating this $M$ times for a fixed $l$ results in an $M$-sample dataset: $\mathcal{K}_l = \left\{\mathcal{T}_l^{(1)}, \mathcal{T}_l^{(2)},..., \mathcal{T}_l^{(M)}  \right\}$. 
In the following, we show that the datasets $\mathcal{K}_1$ and $\mathcal{K}_2$ suffice to bound influences, while including $\mathcal{K}_3$ yields tighter bounds.
\par
In the classical post-processing stage, to estimate the influence on a qubit set $S \subseteq [n]$, we assign to each sample $\mathcal{T}_l^{(j)}$ a binary indicator, \textit{raw influence sampler}, as
\begin{equation}
  X_l^{S}[j] =
  \begin{cases}
    0, & \mathcal{T}_l^{(j)} \cap S = \varnothing,\\
    1, & \mathcal{T}_l^{(j)} \cap S \neq \varnothing.
  \end{cases}
  \label{eq_Xl}    
\end{equation}
Thus $X_l^{S}[j]=1$ when at least one qubit in $S$ is flipped in the $j$-th trial under $U_l$.
We than define the \textit{influence sampler} as the expectation $\mathbb{E}X_l^{S}$, which is estimated by $\overline{X}_l^{S} = \sum_{j=1}^{M}X_l^{S}[j]/M$.
Indeed, $\mathbb{E}X_l^{S}$ is the probability that at least one qubit in $S$ appears in $\T_l$.
Using the first two test gates yields two influence samplers, $\mathbb{E} X^S_1$ and $\mathbb{E} X^S_2$, which already bound the influence $\inf_S[\Phi]$:
\begin{equation}
        \begin{aligned}[b]
\mathrm{IL}_S\coloneqq\max_{l\in\left\{1,2\right\}}\left\{\mathbb{E}X_l^S\right\}\le \, \inf_S[\Phi]
        \le \mathrm{IU}_S \coloneqq \sum_{l\in \left\{1,2\right\}}\mathbb{E}X_l^S.
        \end{aligned}
        \label{ieq_1}
\end{equation}
Including the third influence sampler $\mathbb{E} X^S_3$ tightens these to
\begin{equation}
\adjustbox{scale=0.9}{$
        \begin{aligned}[b]
        \mathrm{IL}_S^{(\mathrm{II})}\coloneqq\max_{l\in\left\{1,2,3\right\}}\left\{\mathbb{E}X_l^S\right\}\le \, \inf_S[\Phi]\le \mathrm{IU}_S^{(\mathrm{II})} \coloneqq \frac{1}{2} \sum_{l\in \left\{1,2,3\right\}}\mathbb{E}X_l^S.
        \end{aligned}
        $}
\label{ieq_2}
\end{equation}
Here, the \textit{influence bounds} $\mathrm{IL}$ and $\mathrm{IU}$ represent the lower and upper bounds, respectively.
As shown in appendix~\hyperref[app_c]{C}, $\text{IL}_S^{(\mathrm{II})}$ and $\text{IU}_S^{(\mathrm{II})}$ are consistently tighter, since three samplers probe more diagonal $\chi$-elements.
In particular, for single-qubit sets $S = \{i\}$, one can find $\text{Inf}_{\{i\}}[\Phi] = \text{IU}^{(\mathrm{II})}_{\{i\}}$, because three test gates generate single-qubit 2-design states~\cite{1523643}, enabling direct estimation of the process fidelity~\cite{PhysRevA.101.042316}.
Indeed, the upper bounds approximate the true influence within constant factors: $ \inf_S[\Phi] \leq \text{IU}_S \leq 2  \inf_S[\Phi]$ and $\inf_S[\Phi] \leq \mathrm{IU}_S^{(\mathrm{II})} \leq 3\inf_S[\Phi]/2$.
\par
\subsection{Identifying high-influence qubits}
With influence sampling, we aim to find the minimal subset $T \subseteq [n]$ such that the $T$-junta $\Phi_T\otimes\mathcal{I}_{T^c}$ approximates $\Phi$ with an error $D(\Phi_T\otimes\mathcal{I}_{T^c},\Phi)\leq \epsilon$, for a given small $\epsilon$. 
By equation~(\ref{eq_main_the}), this requires $\inf_{T^c}[\Phi]\leq \delta$ with $\delta = \left(\sqrt{\sqrt{2}\epsilon+1/2}-1/\sqrt{2}\right)^2$.
\par
To realize this, influence sampling is performed to identify a minimal $T$ for which $\text{IU}_{T^c}$ or $\text{IU}^{(\mathrm{II})}_{T^c}$ is at most $\delta$, using a decision threshold $\delta_0\leq\delta$.
Employing the large deviation principle with a worst-case two-point reduction~\cite{Dembo2010}, we obtain
\begin{equation}
    \Pr(\overline{Y}_i\leq\delta_0,\ \Inf_{T^c}[\Phi]\geq \delta)\leq \exp[- D_{\mathrm{KL}}\left(\frac{\delta_0}{L_i}\Big\Vert \frac{\delta}{L_i} \right)M]
    \label{eq_ldp}
\end{equation}
with $i=1,2$, where $D_{\mathrm{KL}}$ is the Kullback-Leibler divergence. 
Here, $\overline{Y}_1=\sum_{l=1}^2\overline{X}^{T^c}_l$ ($\overline{Y}_2=\sum_{l=1}^3\overline{X}^{T^c}_l$) denotes the estimation of $\text{IU}_{T^c}$ ($\text{IU}^{(\mathrm{II})}_{T^c}$) with the value range length $L_1=2$ ($L_2=3/2$).
Given a confidence level $1-\eta$, equation~(\ref{eq_ldp}) implies that observing $\overline{Y}_i\leq \delta_0$ with $M$ samples suffices to conclude that $D(\Phi_T\otimes\mathcal{I}_{T^c},\Phi)\leq\epsilon$.
As detailed in appendix~\hyperref[app_d]{D}, for small $\epsilon$, this requires a sample complexity of $M=\mathcal{O}(\kappa^2\ln(1/\eta)/\epsilon^2)$ with $\kappa := \delta/(\delta-\delta_0)$.
In practice, we first collect $M$ samples $\T_l$ per $U_l$ during quantum sampling to form the datasets $\{\mathcal{K}_l\}$, and then select any $S$ to compute its influence bounds using this fixed data in classical post-processing.
Thus, varying $S$ after quantum sampling incurs no additional sample complexity. 
\par
In real quantum devices, state preparation and measurement (SPAM) errors~\cite{PhysRevResearch.4.013199}, here taken to include test-gate imperfections, are unavoidable. Thus, even qubits that ideally have zero influence from $\Phi$ may be observed with small but nonzero influence values.
To identify high-influence qubit subset, we set a decision threshold $\delta_0$ on single-qubit influence, and include the qubit $i$ in $T$ when $\mathrm{IU}_{\{i\}}> \delta_0$. 
The measured $\mathrm{IU}_{T^c}$ is then used to bound the junta-approximation error using equation~(\ref{eq_main_the}).
The algorithmic descriptions of influence‐sampling are given in appendix~\hyperref[app_e]{E}.

\section{Experimental setup}
The experimental setup for influence sampling on a four-qubit photonic circuit is shown in figure~\hyperref[fig1_schematic_exp]{1(b)}. 
Photon pairs produced via spontaneous parametric down-conversion in a $\beta$-barium-borate (BBO) crystal, pumped by a frequency-doubled Ti:sapphire laser, are injected into two separate linear-optical circuits. 
Each photon encodes a two-qubit product state using the path and polarization degrees of freedom (DoF), yielding the four-qubit state $\ket{q_1}\ket{q_2}\ket{q_3}\ket{q_4}$.
For the first (third) photon, $\ket{q_1}$ ($\ket{q_3}$) denotes
the path qubit with upper ($s_0$) and lower ($s_1$) paths, and $\ket{q_2}$ ($\ket{q_4}$) denotes the polarization qubit with horizontal ($H$) and vertical ($V$) modes. 
The computational basis $\{\ket{0},\ket{1}\}$ is encoded as $\{\ket{s_0},\ket{s_1}\}$ for path, and $\{\ket{H},\ket{V}\}$ for polarization (details in appendices \hyperref[app_f]{F} and \hyperref[app_g]{G}).
\par
We design four-qubit quantum processes as $k$-junta with $k$ ranging from 0 to 3, realized by combining single-qubit subprocesses $\Phi_{\{i\}}$, non-separable two-qubit subprocesses $\Phi_{\{i,j\}}$, and identities, with subscripts indicating the acted-on qubits.
For example, a 2-junta can be implemented as the forms of $\Phi_{\{1,2\}} \otimes \mathcal{I}_{\{3,4\}}$ or $\Phi_{\{1\}} \otimes\mathcal{I}\otimes
\Phi_{\{3\}} \otimes \mathcal{I}_{\{4\}}$. 
Figure~\hyperref[fig1_schematic_exp]{1(c)} shows the photonic implementations of these subprocesses, including unitary gates and phase-damping channels (details in appendix \hyperref[app_h]{H}).
\par
\begin{figure*}[ht]
    \centering
    \includegraphics[width= 1 \textwidth]{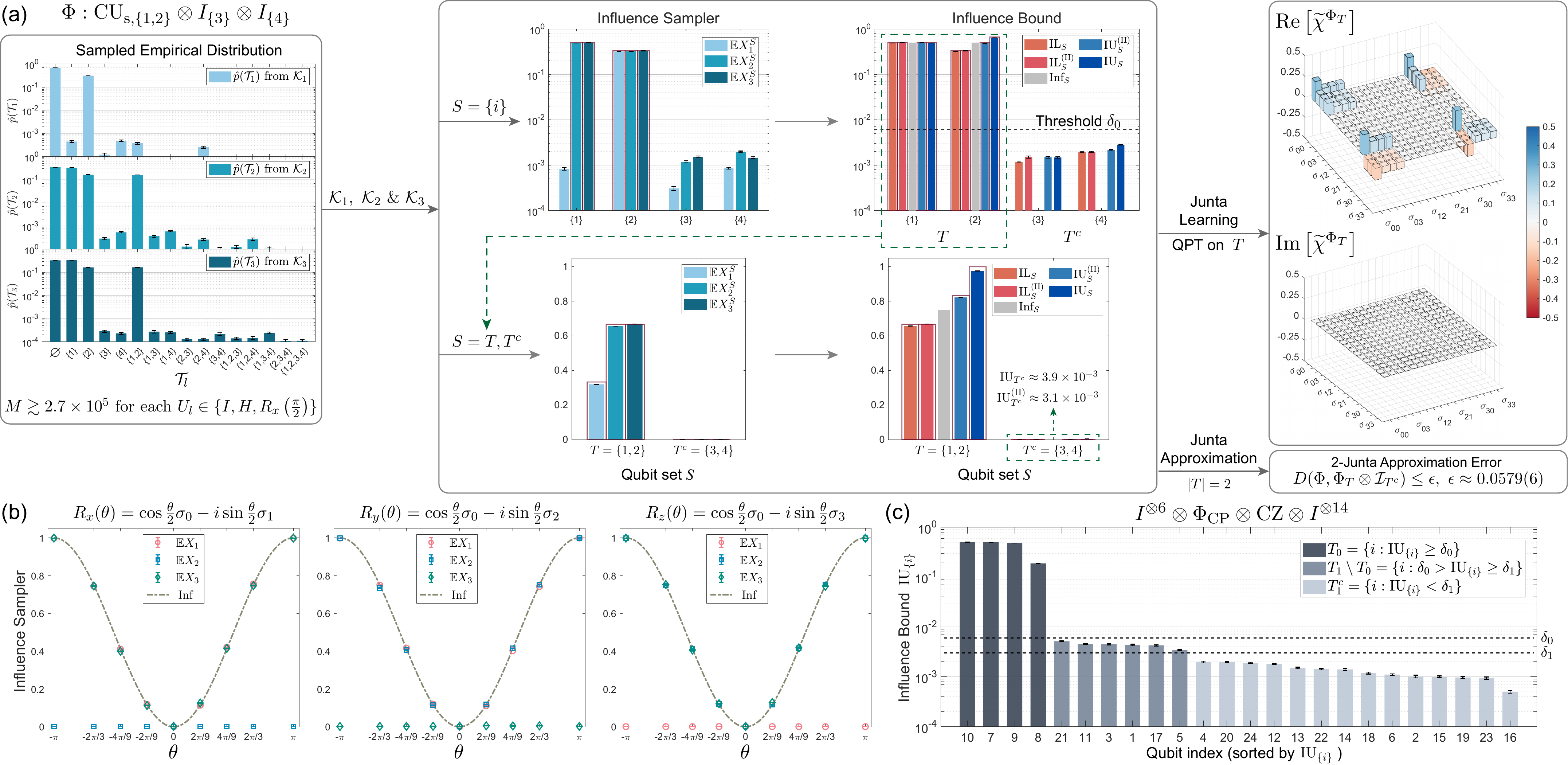}
    \caption{(a) Experimental influence sampling and junta process learning.
    The target quantum process is 
    $\mathrm{CU_{\mathrm{s},\{1,2\}}}\otimes I_{\{3\}}\otimes I_{\{4\}}$.  
    $\mathrm{CU}_{\mathrm{s},\{1,2\}}$ is a controlled-$U_s$ gate with $\ket{q_1}$ as control and $\ket{q_2}$ as target;
    $U_s = \sum_{j=1}^3\sigma_j/\sqrt{3}$.
    For each test gate $U_l$, we acquire $M \gtrsim 2.7 \times 10^5$ samples of $\T_l$, forming the collection $\mathcal{K}_l$.
    The empirical distributions $\hat{p}(\T_l)$ are shown on the left inset.
    From these we compute influence samplers $\mathbb{E}X^S_l$ for various qubit sets $S$ (middle left), and obtain lower ($\text{IL}_S$, $\text{IL}_S^{(\mathrm{II})}$) and upper bounds ($\text{IU}_S$, $\text{IU}_S^{(\mathrm{II})}$) on $\inf_S[\Phi]$ (middle right).
    $\text{IU}^{(\mathrm{II})}_{\{i\}}$ identify the high-influence subset $T$ and its complement $T^c$. 
    $\text{IU}_{T^c}^{(\mathrm{II})}$ yields the bound $\epsilon$ on the junta‑approximation error.
    The subprocess $\Phi_T$ on $T$ is then learned by QPT, giving $\widetilde{\chi}^{\Phi_T}$ (right inset) and hence the 2-junta approximation $\widetilde{\Phi}_T\otimes\mathcal{I}_{T^c}$.
    Red frames outside the bars indicate theoretical values. Gray bars between $\text{IL}_S$ and $\text{IU}_S$ represent theoretical influence values. 
    (b) Measured influence samplers for three single-qubit rotation gates.
    (c) Influence sampling on a 24-qubit quantum circuit, where the process consists primarily of a CZ gate and a controlled phase-damping process.
    The dashed black lines mark the single‑qubit decision thresholds $\delta_0 = 0.006$ and $\delta_1 = 0.003$.
    Error bars are generated via Poisson distribution.    }
    \label{fig2_IF_result_1}
\end{figure*}
Computational basis measurements are performed on each photon using a toggling half-wave plate (HWP) and a beam displacer, followed by two single-photon detection modules (SPCMs). In figure~\hyperref[fig1_schematic_exp]{1(b)}, toggling HWPs are set to $0^\circ$ to measure $\{\ket{00},\ket{01}\}$ and to $45^\circ$ for $\{\ket{10},\ket{11}\}$. 
Coincidence events from the SPCMs are recorded by a field-programmable gate array (FPGA).
Four coincidence channels, corresponding to SPCM pairs $\{13,14,23,24\}$, together with the four settings of TH1 and TH2, implement measurements in the complete four-qubit computational basis (details in appendix \hyperref[app_i]{I}).

\section{Experimental results}
To perform influence sampling, each qubit $\ket{q_i}$ is randomly prepared in $\ket{0}$ or $\ket{1}$ with equal probability and then undergoes a test gate $U_l\in\{I,H,R_x(\frac{\pi}{2})\}$. 
State preparation and gate operation, $U_l\ket{q_i}$, are implemented jointly using a HWP and a removable quarter-wave plate (QWP) for each qubit. 
After passing the four-qubit process $\Phi$, the inverse test gate $U_l^\dagger$ is applied to each qubit using another removable QWP and HWP. 
For each choice of $U_l$, $M$ random preparations followed by computational-basis measurements yield the collection $\mathcal{K}_l = \{\mathcal{T}_l^{(j)}\}_{j=1}^M$ of sampled high-influence subsets.
\par
Figure~\hyperref[fig2_IF_result_1]{2(a)} presents the experimental results of influence sampling on $\Phi$, which consists of a controlled-$U_s$ gate acting on $\ket{q_1}\ket{q_2}$ and an identity on $\ket{q_3}\ket{q_4}$, with $U_s = \sum_{j=1}^3\sigma_j/\sqrt{3}$.
The left inset shows the empirical distributions of $\T_l$ obtained from the corresponding $\mathcal{K}_l$, collected in the quantum sampling stage with sufficient shots, $M\gtrsim 2.7\times 10^{5}$ for each $U_l$.
This sampling budget suppresses the statistical errors; for example, the estimated influence sampler achieves a $2.8\%$ standard deviation when the true value is $0.005$.
We therefore concentrate on experimentally demonstrating the effectiveness of influence sampling and the junta approximation of quantum processes, rather than on sampling statistical fluctuations.
The middle inset of figure~\hyperref[fig2_IF_result_1]{2(a)} illustrates the classical post-processing to identify high-influence qubits.
We firstly treat each qubit as a target set ($S=\{i\}$) and calculate influence samplers $\mathbb{E}X_l^{\{i\}}$ via equation ~(\ref{eq_Xl}), then yielding influence bounds for each qubit.
The threshold $\delta_0 = 0.006$ is set for single qubits to tolerate SPAM errors in our setup, while details of the error analysis are provided in appendix~\hyperref[app_j]{J}. 
Qubits whose influence upper bounds exceed $\delta_0$ are assigned to the high‑influence subset $T$, with the remaining qubits in its complement $T^{c}$.
We then take $S\in\{T,T^c\}$ to produce influence bounds on on these sets.
Next, we approximate $\Phi$ by the $T$-junta $\Phi_T\otimes\mathcal{I}_{T^c}$, with the distance bound $\epsilon$ determined by $\mathrm{IU}_{T^c}$ or $\mathrm{IU}^{(\mathrm{II})}_{T^c}$.
To learn the $T$-junta process, we perform the QPT on the subprocess $\Phi_T$ to obtain $\widetilde{\chi}^{\Phi_T}$ (right inset of figure~\hyperref[fig2_IF_result_1]{2(a)}), where qubits in $T^c$ is initialized in the maximally mixed state and traced out at the end.
The process fidelity between the reconstructed $\widetilde{\Phi}_{T}$ and the ideal $\mathrm{CU}_{\mathrm{s}}$ gate is $98.40\%$.
\par 
Our results show that the influence bounds computed using $\{\mathcal{K}_1,\mathcal{K}_2, \mathcal{K}_3\}$---namely, $\mathrm{IL}^{(\mathrm{II})}_S$ and $\mathrm{IU}^{(\mathrm{II})}_S$---are consistently tighter than $\mathrm{IL}_S$ and $\mathrm{IU}_S$ obtained from $\{\mathcal{K}_1,\mathcal{K}_2\}$.
The measured influence upper bounds on the $T^c$ are $\mathrm{IU}_{T^c}=3.90(9)\times 10^{-3}$ and $\mathrm{IU}^{(\mathrm{II})}_{T^c}=3.10(6)\times 10^{-3}$ in figure~\hyperref[fig2_IF_result_1]{2(a)}, with the latter giving a tighter bound $\epsilon=0.0579(6)$ on the junta-approximation error via equation~(\ref{eq_main_the}).
Figure~\hyperref[fig2_IF_result_1]{2(b)} shows the three measured influence samplers for single-qubit rotation gates as functions of the rotation angle $\theta$, revealing that each single sampler omits certain diagonal entries of the $\chi$-matrix. For example, $\mathbb{E}X_1$ cannot capture the influence of $R_z$.
Thus, employing three influence samplers yields tighter bounds than using only two.
Additional influence‐sampling results for various quantum processes are shown in appendix~\hyperref[app_k1]{K.1}.
\par
We further demonstrate influence sampling on a 24-qubit circuit by time-multiplexing our four‑qubit device, with details in appendix~\hyperref[app_k2]{K.2}.
The measured $\mathrm{IU}_{\{i\}}$, sorted by their values, are shown in figure~\hyperref[fig2_IF_result_1]{2(c)}.
Using two thresholds, $\delta_0=0.006$ and $\delta_1=0.003$, we obtain two choices of the high-influence subset, $T_0$ and $T_1$. 
From the measured empirical distributions, we generate $10^5$ Monte Carlo samples to estimate $\mathrm{IU}_{T^c_0} = 0.0424(8)$ and $\mathrm{IU}_{T^c_1} = 0.0178(6)$.
The bounds on junta-approximation error are given by $\epsilon_0 = 0.236(3)$ and $\epsilon_1 = 0.146(3)$.
Given that $|T_0| = 4$ and $|T_1| = 10$, these results illustrate the trade‑off between a smaller junta size and a smaller approximation error. 
\par
\begin{figure}[ht]
    \centering
    \includegraphics[width=0.5\textwidth]{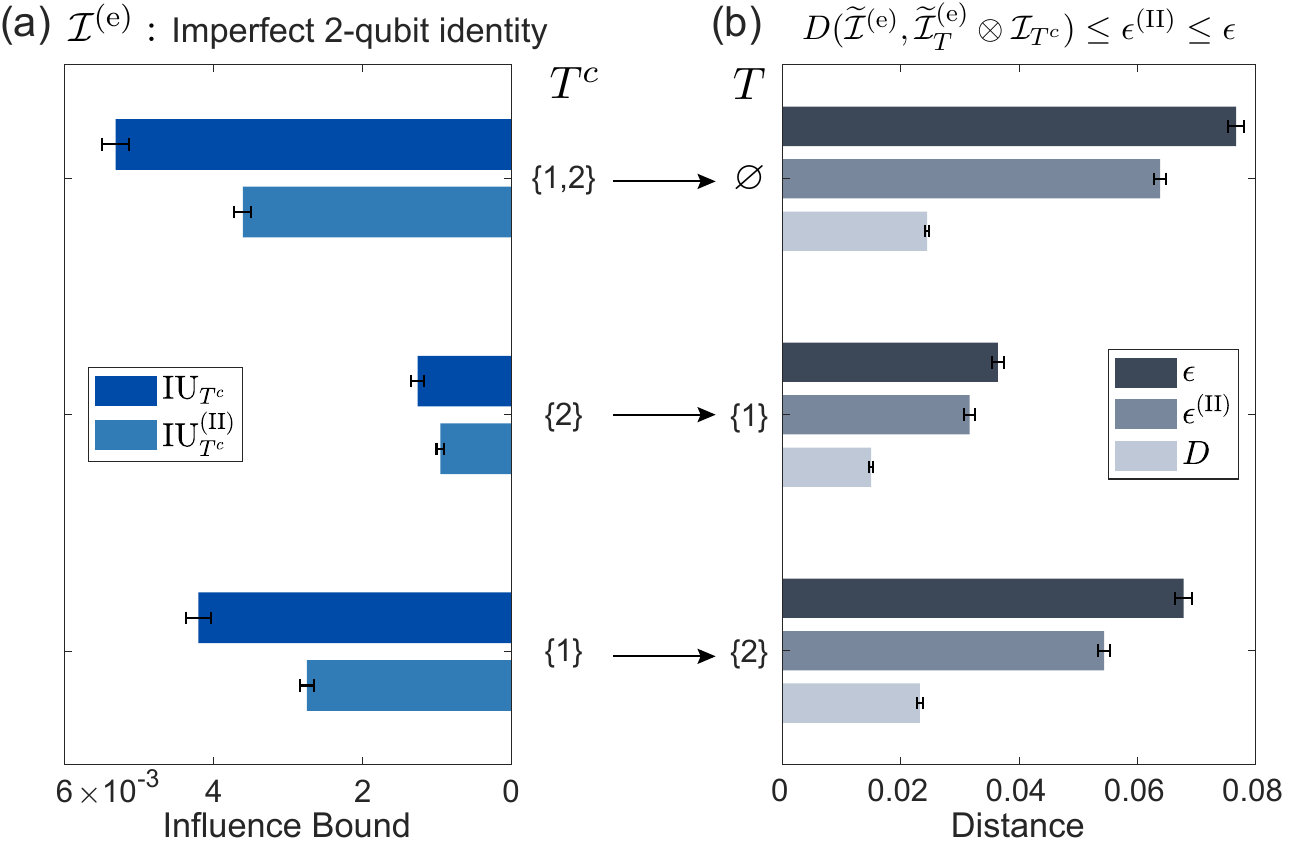}
    \caption{(a) Measured influence bounds for an imperfect two-qubit identity process $\mathcal{I}^{\mathrm{(e)}}$ with $M\approx1.4\times10^5$. 
    (b) Experimental evaluation of the junta-approximation distances and their bounds. $\widetilde{\mathcal{I}}^{\mathrm{(e)}}$ and $\widetilde{\mathcal{I}}^{\mathrm{(e)}}_{T}$ are reconstructed via two-qubit QPT, and via single-qubit QPT with the qubits in $T^c$ initialized in the maximally mixed state and subsequently traced out, respectively.
    }
    \label{fig3_DB_result_1}
\end{figure}

To validate the junta-approximation error bound, we first preform influence sampling on an imperfect two-qubit identity process $\mathcal{I}^{\mathrm{(e)}}$, thereby obtaining influence bounds for different choices of $T^c\subseteq \{1,2\}$, as shown in figure~\hyperref[fig3_DB_result_1]{3(a)}. 
For each choice of $T$ and $T^c$, we then perform junta learning to reconstruct the subprocess on $T$ as $\widetilde{\mathcal{I}}^{\mathrm{(e)}}_T$. 
In parallel, full two-qubit QPT on the whole $\mathcal{I}^{\mathrm{(e)}}$ yields a reconstruction $\widetilde{\mathcal{I}}^{\mathrm{(e)}}$. 
The junta-approximation distance is then estimated as 
$D(\widetilde{\mathcal{I}}^{\mathrm{(e)}},\widetilde{\mathcal{I}}^{\mathrm{(e)}}_T\otimes\mathcal{I}_{T^c})$, with the corresponding bounds $\epsilon$ and $\epsilon^{\mathrm{(II)}}$ derived from the measured $\mathrm{IU}_{T^c}$ and $\mathrm{IU}_{T^c}^{\mathrm{(II)}}$.
As shown in figure~\hyperref[fig3_DB_result_1]{3(b)}, the estimated distance lies well within these bounds, thereby demonstrating the effectiveness of the junta approximation enabled by influence sampling. 
Further details are in appendix~\hyperref[app_k3]{K.3}.

\section{Conclusions}
In this paper, we present an influence sampling protocol for characterizing $n$‑qubit quantum processes. Using only two or three  test gates, it estimates valid influence bounds over all $2^n$ qubit subsets with sample complexity independent of $n$. Compared with the Influence-Estimator approach in \cite{Thomas2023Testing}, it is more hardware‑friendly, requiring neither entangled state preparation nor Bell measurements.
Leveraging a photonic platform, we implement this protocol on several processes and observe the influence bounds that agree with theory and cleanly separate influenced from uninfluenced qubits; using three test gates further tightens the bounds relative to two gates, as confirmed experimentally.
We apply influence sampling to a 24‑qubit circuit, demonstrating scalability and the size–error trade‑off of the junta approximation, and we further validate the approximation on an imperfect two‑qubit identity process by comparing the estimated distances with the estimated bounds.
\par
With low hardware overhead and sample complexity independent of system size, influence sampling provides a practical addition to the toolbox of quantum certification and benchmarking protocols~\cite{RN96,Boixo2018,PhysRevLett.122.210502,PhysRevLett.124.010504}. 
It can be naturally combined with existing techniques, including tomography~\cite{PhysRevA.87.062119,Torlai2023,PhysRevA.101.022317}, verification~\cite{PhysRevLett.114.140505,PhysRevLett.128.020502}, and randomized benchmarking~\cite{PhysRevLett.108.260503,PhysRevLett.109.080505}, yielding substantial speedups when the underlying process is close to a junta~\cite{bao2023testing}. 
We also expect that it can be used to address quantum gates, identify qubits highly affected by SPAM errors, and detect gate crosstalk~\cite{PRXQuantum.3.020301}.
These results demonstrate influence sampling as an effective and scalable approach for extracting key structural information from multi‑qubit quantum processes.

\section*{Acknowledgements}
All authors acknowledge support from the National Natural Science Foundation of China (Grants No. 12347104, 12461160276) and the Natural Science Foundation of Jiangsu Province (Grants No. BK20243060).
H.Z., Q.W., M.M., and L.Z. also acknowledge support from the National Key Research and Development Program of China (Grants No. 2023YFC2205802), the National Natural Science Foundation of China (No. U24A2017), the Natural Science Foundation of Jiangsu Province (Grants No. BK20233001), and in part by State Key Laboratory of Advanced Optical Communication Systems and Networks, China. 
Z.B., Z.Y. and P.Y. also acknowledge support from the National Natural Science Foundation of China (Grant No. 62332009), the Innovation Program for Quantum Science and Technology (Grant No. 2021ZD0302901) and the New Cornerstone Science Foundation.
This work was done when Z.B. was a master student at Nanjing University.

\section*{Data availability}
The data supporting the plots and other findings in this paper are openly available. 
The dataset is accessible at \url{https://doi.org/10.5281/zenodo.18244131}.

\bibliographystyle{quantum}
\bibliography{references}

\onecolumn
\appendix
\section{Theoretical preliminaries}
\label{app_a}
Consider an $n$-qubit quantum system with Hilbert space $\mathcal{H}$. Let $\mathcal{L}(\mathcal{H})$ denote the space of linear operators on $\mathcal{H}$. 
A quantum process $\Phi$ is a completely positive and trace-preserving (CPTP) map from $\mathcal{L}(\mathcal{H})$ to $\mathcal{L}(\mathcal{H})$. 
For a set of qubits $S\subseteq [n]$, where qubits are represented by their corresponding subscripts, we define an $S$-junta process as $\Phi = \Phi_S \otimes \mathcal{I}_{S^c}$. 
Here, $\Phi_S$ acts non-trivially only on qubits in $S$, while $\mathcal{I}$ represents the identity process on the complement $S^c = [n]\backslash S$. If $|S| \le k$, then $\Phi$ is also called a $k$-junta process.
Quantum processes can be represented in several ways, including the \textit{Kraus}, \textit{Choi}, and \textit{process matrix} representations~\cite{wood2015tensor}. 
\par
\textbf{Kraus representation.} The Kraus representation of the quantum process $\Phi$ is given by:
\begin{equation}
    \Phi(\rho) = \sum_{i=1}^m K_i\rho K_i^{\dagger},
\end{equation}
where $\rho\in \mathcal{L}(\mathcal{H})$ is the density matrix, and $\{K_i\}_{i=1}^{m}$ are Kraus operators satisfying the CPTP condition $\sum_{i}K_i^{\dagger}K_i=I$ with $I$ representing the identity operator on $\mathcal{H}$. 
\par
\textbf{Choi representation.} The (unnormalized) Choi representation of $\Phi$ is given by~\cite{CHOI1975285}:
\begin{equation}
    J^{\Phi} = (\Phi\otimes I)\left(\sum_{a,b= 0}^{d-1}\ketbra{aa}{bb}\right) =\sum_{a,b=0}^{d-1}\Phi(\ketbra{a}{b})\otimes \ketbra{a}{b},
\end{equation}
where $\text{Tr}(J^{\Phi}) = d$ and $d=2^n$ is the dimension of the $n$-qubit system. 
The inner product between two quantum processes $\Phi$ and $\Psi$ is defined as the Hilbert-Schmidt inner product between their Choi representations, i.e., $\langle\Psi,\Phi\rangle = \langle J^{\Psi}, J^{\Phi}\rangle = \text{Tr}({J^{\Psi}}^\dagger J^{\Phi})$.  
The \textit{vectorization} of matrices, denoted by $\text{vec}(\cdot)$, transforms a $m\times m$ matrix to a $m^2\times 1$ column vector. 
For the standard basis $\ketbra{a}{b}$ of $\mathcal{L}(\mathcal{H})$, the \textit{row-stacking} vectorization is defined as $\text{vec}(\ketbra{a}{b}) = \ket{a}\!\ket{b}$~\cite{wood2015tensor}. Consequently, the Choi matrix can then be expressed in terms of the Kraus operators as:
\begin{equation}
    J^{\Phi} = \sum_{i=1}^m \text{vec}(K_i)\text{vec}(K_i)^\dagger.
\end{equation}
\par
\textbf{Process matrix representation.}
Note that the Fourier expansion of $\Phi$ in~\cite{bao2023testing} is equivalent to the process matrix (also known as the $\chi$-matrix) representation. The process matrix representation of $\Phi$ is given by~\cite{Nielsen_Chuang_2010} 
\begin{equation}
    \Phi(\rho) = \sum_{\mathbf{x},\mathbf{y}\in \mathbb{Z}_4^{n}}\chi^{\Phi}_{\mathbf{x}\mathbf{y}}\sigma_\mathbf{x}\rho\sigma_\mathbf{y},
\end{equation}
where $\chi^{\Phi}_{\mathbf{x}\mathbf{y}}$ are the non-negative $\chi$-matrix coefficients of $\Phi$, satisfying $\chi^{\Phi}_{\mathbf{x}\mathbf{y}} = \langle \Phi_{\mathbf{x},\mathbf{y}},\Phi \rangle/d^2$ with $\Phi_{\mathbf{x},\mathbf{y}}(\rho) = \sigma_\mathbf{x}\rho\sigma_\mathbf{y}$. Let $\mathbb{Z}_4=\{0,1,2,3\}$ and $\mathbb{Z}_4^{n} = \{0,1,2,3\}^{n}$. 
Moreover, $\sigma_\mathbf{x} =\bigotimes_{i = 1}^{n} \sigma_{x_i}$ and $\sigma_\mathbf{y} =\bigotimes_{i = 1}^{n} \sigma_{y_i}$ are the general multi-qubit Pauli operators, where $x_i, y_i \in \mathbb{Z}_4$, and $\mathbf{x}=(x_1,x_2,...,x_n)\in \mathbb{Z}_4^{n}$, $\mathbf{y}=(y_1,y_2,...,y_n)\in \mathbb{Z}_4^{n}$. $\{\sigma_x\}_{x=0}^{3}$ denote identity and three single-qubit Pauli operators:
\begin{equation}
    \sigma_0 = \left( \begin{matrix} 1 & 0\\ 0 & 1\end{matrix}\right)=I,\ \ \sigma_1 = \left( \begin{matrix} 0 & 1\\ 1 & 0\end{matrix}\right)=X,\ \ 
    \sigma_2 = \left( \begin{matrix} 0 & -i\\ i & 0\end{matrix}\right)=Y,\ \ 
    \sigma_3 = \left( \begin{matrix} 1 & 0\\ 0 & -1\end{matrix}\right)=Z.\ \ 
    \label{eq_pauli}
\end{equation}
The process matrix $\chi^{\Phi} = [\chi^{\Phi}_{\mathbf{x}\mathbf{y}}]$ is commonly used in quantum process tomography. Both Choi matrix $J^{\Phi}$ and process matrix $\chi^{\Phi}$ are positive and Hermitian matrices, and are related by a unitary transformation $U$: 
\begin{equation}
    \chi^{\Phi} = \frac{1}{d}U^{\dagger}J^{\Phi}U.
\end{equation}
The unitary $U$ can be expressed as the ordered combination of the vectorization of the multi-qubit Pauli operators as $U = \left[ \text{vec}(\sigma_\mathbf{x})/\sqrt{d} \right]_{\mathbf{x}\in \mathbb{Z}_4^{n}}$.
\par
We further introduce the distance and fidelity between two quantum processes, define the reduced quantum subprocess, and then present the definition and intuition of the influence of quantum processes.
\par
\textbf{Process distance.} The distance between two quantum processes, $\Phi$ and $\Psi$, is defined as
\begin{equation}
    D\left(\Phi,\Psi\right) = \frac{1}{\sqrt{2}}\Vert \chi^{\Phi}-\chi^{\Psi}\Vert_F = \frac{1}{\sqrt{2}d}\Vert J^{\Phi}-J^{\Psi}\Vert_F ,
\end{equation}
where $\Vert\cdot\Vert_F$ denotes the Frobenius norm, given by $\Vert A \Vert_F = \sqrt{\text{Tr}(A^{\dagger}A)}$ and the normalized factor $1/\sqrt{2}d$ ensures that $D(\Phi,\Psi)\in[0,1]$. 
For two unitary processes $\mathcal{U}$ and $\mathcal{V}$, the distance is given by $D(\mathcal{U},\mathcal{V})=\sqrt{1-\Tr(\chi^{\mathcal{U}}\chi^{\mathcal{V}})}=\sqrt{1-\Tr(J^{\mathcal{U}}J^{\mathcal{V}})/d^2}$.
Further discussion on the properties of this distance function and its relationship to other metrics, particularly its average-case characteristic, can be found in \cite{bao2023testing}. 
We highlight that there might be essentially no efficient algorithm for the property testing problem equipped with the worst-case operator norm. For more details, interested readers are referred to Section 5.1.1 of~\cite{montanaro2018survey}.
\par
\textbf{Process fidelity.} The process fidelity is defined as 
\begin{equation}
    F(\Phi,\Psi) = \left(\text{Tr}\sqrt{\sqrt{\chi^{\Phi}}\chi^{\Psi}\sqrt{\chi^{\Phi}}}\right)^2 = \frac{1}{d^2}\left(\text{Tr}\sqrt{\sqrt{J^{\Phi}}J^{\Psi}\sqrt{J^{\Phi}}}\right)^2 .
\end{equation}
When the quantum process $\Psi$ is a unitary process $\mathcal{U}$, its Choi representation is rank-$1$, and the process fidelity is equal to the entanglement gate fidelity: $F_e(\Phi,\mathcal{U}) = \text{Tr}(\chi^{\Phi}\chi^{\mathcal{U}}) = \text{Tr}(J^{\Phi}J^{\mathcal{U}})/d^2 $~\cite{PhysRevA.101.042316}.
\par
\textbf{Reduced quantum subprocess.} 
To describe the action of $\Phi$ on a qubit subset $S$, we define an reduced quantum subprocess $\Phi_S$ acting on $S$ by
\begin{equation}
\begin{split}
    \Phi_S(\rho_S) & = \mathrm{Tr}_{S^c}\left[\Phi\left(\rho_S\otimes \frac{I_{S^c}}{2^{|S^c|}}\right)\right] = \sum_{\mathbf{x},\mathbf{y}\in\Zfn}\chi^{\Phi}_{\mathbf{x}\mathbf{y}}\Tr_{S^c}\lrb{\sigma_{\mathbf{x}_S}\rho_S\sigma_{\mathbf{y}_S}\otimes \f{\sigma_{\mathbf{x}_{S^c}}\sigma_{\mathbf{y}_{S^c}}}{2^{\abs{S^c}}}} = \sum_{\mathbf{x},\mathbf{y}\in \Zfn\atop \mathbf{x}_{S^c}=\mathbf{y}_{S^c}} \chi^{\Phi}_{\mathbf{x}\mathbf{y}}\sigma_{\mathbf{x}_S}\rho_S\sigma_{\mathbf{y}_S} \\
    & = \sum_{\mathbf{x}_S,\mathbf{y}_S\in \mathbb{Z}_4^{|S|}} \chi^{\Phi_S}_{\mathbf{x_S}\mathbf{y_S}}\sigma_{\mathbf{x}_S}\rho_S\sigma_{\mathbf{y}_S},
\end{split}
\end{equation}
where $\chi^{\Phi_S}_{\mathbf{x_S}\mathbf{y_S}} = \sum_{\mathbf{z}_{S^c}\in \mathbb{Z}_4^{|S^c|}} \chi^{\Phi}_{(\mathbf{x}_S\circ\mathbf{z}_{S^c},\mathbf{y}_S\circ\mathbf{z}_{S^c})}$. Here, $\circ$ denotes concatenation, $\mathbf{x}_S$ and $\mathbf{x}_{S^c}$ represent the index vectors on subsets $S$ and $S^c$ respectively, and $\chi^{\Phi_S}$ is the process matrix of $\Phi_S$. 
Operationally, $\Phi_S$ is obtained by applying $\Phi$ to the input $\rho_S \otimes I_{S^c}/2^{|S^c|}$ (the maximally mixed state on $S^c$) and tracing out the qubits in $S^c$.
\par
\textbf{Influence.} 
To qualify the impact of the quantum process $\Phi$ on a qubit subset $S$, define the influence of $\Phi$ on $S$ as
\begin{equation}
    \text{Inf}_{S}[\Phi] = \sum_{\exists i\in S, x_i \neq 0}\chi^{\Phi}_{\mathbf{x}\mathbf{x}}=1-\sum_{\substack{\forall i\in S,\  x_i=0 } }\chi^{\Phi}_{\mathbf{x}\mathbf{x}},
\end{equation}
where $\chi^{\Phi}_{\mathbf{x}\mathbf{x}}$ is the diagonal elements of the process matrix.
This measure captures the contribution of the diagonal elements corresponding to non-trivial Pauli operators acting on the qubits in $S$.
\par
The process fidelity between the reduced subprocess $\Phi_S$ and the identity process $\mathcal{I}_S$ is:
\begin{equation}
    F(\Phi_S,\mathcal{I}_S)=\text{Tr}(\chi^{\Phi_S}\chi^{\mathcal{I}_S})=\chi^{\Phi_S}_{\mathbf{0}\mathbf{0}} = \sum_{\mathbf{z}_{S^c}\in \mathbb{Z}_4^{|S^c|}} \chi^{\Phi}_{(\mathbf{0}_S\circ\mathbf{z}_{S^c},\mathbf{0}_S\circ\mathbf{z}_{S^c})}=\sum_{\substack{\forall i\in S,\  x_i=0 } }\chi^{\Phi}_{\mathbf{x}\mathbf{x}}=1-\inf_S[\Phi].
\end{equation}
Therefore, the influence can be expressed as the infidelity of $\Phi_S$ as $\inf_S[\Phi] = 1 - F(\Phi_S,\mathcal{I}_S)$.
\par
On the other hand, influence quantifies the magnitude of process action. Consider an $n$-qubit unitary process $\mathcal{U}(\cdot) = U(\cdot)U^{\dagger}$ with $U$ can be decomposed as
\begin{equation}
    U = e^{-i\theta H} = (I\cos{\theta}-iH\sin{\theta}),
\end{equation}
where $H$ is a traceless Hermitian generator satisfying $H^2=I$. 
The generator can be expressed as $H = \sum_{\mathbf{x}\neq \mathbf{0}}\alpha_\mathbf{x} \sigma_\mathbf{x}$, where the coefficients $\alpha_\mathbf{x}$ are real and normalized such that $\sum_{\mathbf{x}\neq\mathbf{0}}\alpha_{\mathbf{x}}^2 = 1$. 
The Pauli-operator expansion of the process $\mathcal{U}$ is given by
\begin{equation}
    \mathcal{U}(\rho) = \cos^2{\theta}\cdot\sigma_\mathbf{0}\rho\sigma_\mathbf{0}+i\cos{\theta}\sin{\theta}\sum_{\mathbf{x}\neq\mathbf{0}}\alpha_{\mathbf{x}}\sigma_{\mathbf{0}}\rho\sigma_{\mathbf{x}}-i\cos{\theta}\sin{\theta}\sum_{\mathbf{x}\neq\mathbf{0}}\alpha_{\mathbf{x}}\sigma_{\mathbf{x}}\rho\sigma_{\mathbf{0}}+\sin^2{\theta}\sum_{\mathbf{x}\neq\mathbf{0},\mathbf{y}\neq\mathbf{0}}\alpha_\mathbf{x}\alpha_\mathbf{y}\sigma_\mathbf{x}\rho\sigma_\mathbf{y}.
\end{equation}
The influence of $\mathcal{U}$ on all $n$ qubits is then given by $\Inf_{[n]}[\mathcal{U}] = \sin^2{\theta}\sum_{\mathbf{x}\neq\mathbf{0}}\alpha_{\mathbf{x}}^2 = \sin^2{\theta}$. 
Therefore, the influence of a unitary process directly reflects the magnitude of the transformation applied to the quantum state. 
In the single-qubit case, the influence for a unitary quantifies the magnitude of the rotation on the Bloch sphere.
\section{Bounds on the junta-approximation error via influence}
In this section, we first formalize the upper bound on junta-approximation error in equation~(\ref{theo_main}) as a theorem and provide its proof.
\label{app_b}
\begin{theorem}
    Let $\Phi$ be a quantum process over $n$ qubits and let $T\subseteq [n]$ with complement $T^c$. Define the reduced subprocess on $T$ by $\Phi_T(\rho_T):=\Tr_{T^c}\lrb{\Phi\p{\rho_T\otimes \f{I_{T^c}}{2^{\abs{T^c}}}}}$. Then, $D\left(\Phi, \Phi_T\otimes\mathcal{I}_{T^c}\right)\le \sqrt{\inf_{T^c}[\Phi]} + \f{\sqrt{2}}{2}\inf_{T^c}[\Phi]$.
    \label{theo_main}
\end{theorem}
\begin{proof}
    For $\mathbf{x},\mathbf{y}\in \Zfn$, we have the process matrix representation of $\Phi$ as $
        \Phi = \sum_{\mathbf{x},\mathbf{y}\in \mathbb{Z}_4^{n}}\chi^{\Phi}_{\mathbf{x}\mathbf{y}}\sigma_\mathbf{x}\rho\sigma_\mathbf{y}.$
    The process matrix representation of the reduced subprocess $\Phi_T$ is given by:
    \begin{align*}
        \Phi_T(\rho_T) &= \Tr_{T^c}\lrb{\Phi\p{\rho_T\otimes \f{I_{T^c}}{2^{\abs{T^c}}}}} \\
        &= \sum_{\mathbf{x},\mathbf{y}\in\Zfn}\chi^{\Phi}_{\mathbf{x}\mathbf{y}}\Tr_{T^c}\lrb{\sigma_{\mathbf{x}_T}\rho_T\sigma_{\mathbf{y}_T}\otimes \f{\sigma_{\mathbf{x}_{T^c}}\sigma_{\mathbf{y}_{T^c}}}{2^{\abs{T^c}}}} \\
        &= \sum_{\mathbf{x},\mathbf{y}\in \Zfn\atop \mathbf{x}_{T^c}=\mathbf{y}_{T^c}} \chi^{\Phi}_{\mathbf{x}\mathbf{y}}\sigma_{\mathbf{x}_T}\rho_T\sigma_{\mathbf{y}_T},
    \end{align*}
    where $\sigma_{\mathbf{x}}=\sigma_{\mathbf{x}_T}\otimes \sigma_{\mathbf{x}_{T^c}}$ and $\sigma_{\mathbf{y}}=\sigma_{\mathbf{y}_T}\otimes \sigma_{\mathbf{y}_{T^c}}$. Therefore, the distance between $\Phi$ and $\Phi_T\otimes\mathcal{I}_{T^c}$ is given by
    \begin{equation}
    \begin{split}
        &\ \ \ \ \frac{1}{\sqrt{2}d}\norm{J(\Phi)-J(\Phi_T\otimes \mathcal{I}_{T^c})}_F \\
        &= 
            \frac{1}{\sqrt{2}d}\norm*{\sum_{\mathbf{x},\mathbf{y}\in\Zfn}\chi^{\Phi}_{\mathbf{x}\mathbf{y}}\vect{\sigma_\mathbf{x}}\vect{\sigma_\mathbf{y}}^{\dagger}
                - \sum_{\mathbf{x},\mathbf{y}\in\Zfn\atop \mathbf{x}_{T^c}=\mathbf{y}_{T^c}} \chi^{\Phi}_{\mathbf{x}\mathbf{y}}\vect{\sigma_{\mathbf{x}_T\circ \mathbf{0}_{T^c}}}\vect{\sigma_{\mathbf{y}_T\circ \mathbf{0}_{T^c}}}^{\dagger} 
            }_F \\
        &= \frac{1}{\sqrt{2}d}\norm*{
            \sum_{\mathbf{x},\mathbf{y}\in\Zfn\atop \mathbf{x}_{T^c}\neq \mathbf{0}_{T^c}\ \mathrm{or} \ \mathbf{y}_{T^c}\neq \mathbf{0}_{T_c}}\chi^{\Phi}_{\mathbf{x}\mathbf{y}}\vect{\sigma_\mathbf{x}}\vect{\sigma_\mathbf{y}}^{\dagger}
            - \sum_{\mathbf{x},\mathbf{y}\in\Zfn\atop \mathbf{x}_{T^c}=\mathbf{y}_{T^c}\neq \mathbf{0}_{T^c}} \chi^{\Phi}_{\mathbf{x}\mathbf{y}}\vect{\sigma_{\mathbf{x}_T\circ \mathbf{0}_{T^c}}}\vect{\sigma_{\mathbf{y}_T\circ \mathbf{0}_{T^c}}}^{\dagger} 
        }_F \\
        &\le \frac{1}{\sqrt{2}d}\norm*{ \sum_{\mathbf{x},\mathbf{y}\in\Zfn\atop \mathbf{x}_{T^c}\neq \mathbf{0}_{T^c}\ \mathrm{or}\ \mathbf{y}_{T^c}\neq \mathbf{0}_{T^c}}\chi^{\Phi}_{\mathbf{x}\mathbf{y}}\vect{\sigma_\mathbf{x}}\vect{\sigma_\mathbf{y}}^{\dagger} }_F
            + \frac{1}{\sqrt{2}d}\norm*{\sum_{\mathbf{x},\mathbf{y}\in\Zfn\atop \mathbf{x}_{T^c}=\mathbf{y}_{T^c}\neq \mathbf{0}_{T^c}} \chi^{\Phi}_{\mathbf{x}\mathbf{y}}\vect{\sigma_{\mathbf{x}_T\circ \mathbf{0}_{T^c}}}\vect{\sigma_{\mathbf{y}_T\circ \mathbf{0}_{T^c}}}^{\dagger} }_F,
    \end{split}
    \label{proof_distance}
    \end{equation}
    with $d=2^n$. Consider the square of the first term on the right-hand side of equation(\ref{proof_distance}), we have 
    \begin{align*}
       \f{1}{2d^2}\norm*{ \sum_{\mathbf{x},\mathbf{y}\in\Zfn\atop \mathbf{x}_{T^c}\neq \mathbf{0}_{T^c}\ \mathrm{or}\  \mathbf{y}_{T^c}\neq \mathbf{0}_{T_c}}\chi^{\Phi}_{\mathbf{x}\mathbf{y}}\vect{\sigma_\mathbf{x}}\vect{\sigma_\mathbf{y}}^{\dagger} }_F^2 &= \frac12\sum_{\mathbf{x},\mathbf{y}\in\Zfn\atop \mathbf{x}_{T^c}\neq \mathbf{0}_{T^c}\ \mathrm{or}\ \mathbf{y}_{T^c}\neq \mathbf{0}_{T_c}}\abs*{\chi^{\Phi}_{\mathbf{x}\mathbf{y}}}^2 \\
       &\le \frac12\sum_{\mathbf{x},\mathbf{y}\in\Zfn\atop \mathbf{x}_{T^c}\neq \mathbf{0}_{T^c}\ \mathrm{or}\  \mathbf{y}_{T^c}\neq \mathbf{0}_{T_c}} \chi^{\Phi}_{\mathbf{x}\mathbf{x}}\chi^{\Phi}_{\mathbf{y}\mathbf{y}} \\
       &\le \frac12\p{\sum_{\mathbf{x},\mathbf{y}\in\Zfn:\  \mathbf{x}_{T^c}\neq \mathbf{0}_{T^c}}+\sum_{\mathbf{x},\mathbf{y}\in\Zfn:\  \mathbf{y}_{T^c}\neq \mathbf{0}_{T^c}}}\chi^{\Phi}_{\mathbf{x}\mathbf{x}}\chi^{\Phi}_{\mathbf{y}\mathbf{y}} \\
       &= \inf_{T^c}[\Phi],
    \end{align*}
    where the first inequality follows that $\chi^{\Phi}$ is positive semi-definite (PSD). 
    \par
    Consider the square of the second term on the right-hand side of equation(\ref{proof_distance}), we have
    \begin{align*}
      &\ \ \ \ \f{1}{2d^2}\norm*{\sum_{\mathbf{x},\mathbf{y}\in\Zfn\atop \mathbf{x}_{T^c}=\mathbf{y}_{T^c}\neq \mathbf{0}_{T^c}} \chi^{\Phi}_{\mathbf{x}\mathbf{y}}\vect{\sigma_{\mathbf{x}_T\circ \mathbf{0}_{T^c}}}\vect{\sigma_{\mathbf{y}_T\circ \mathbf{0}_{T^c}}}^{\dagger} }_F^2 \\
      & = \frac12\sum_{\mathbf{x}_T,\mathbf{y}_T\in \ZfT} \abs*{\sum_{\mathbf{z}_{T^c}\in \ZfTc\atop \mathbf{z}_{T^c}\neq \mathbf{0}_{T^c}}\chi^{\Phi}_{\p{\mathbf{x}_T\circ \mathbf{z}_{T^c}, \mathbf{y}_T\circ \mathbf{z}_{T^c}}}}^2 \\
      & \le \frac12\sum_{\mathbf{x}_T,\mathbf{y}_T\in \ZfT}\p{\sum_{\mathbf{z}_{T^c}\in \ZfTc\atop \mathbf{z}_{T^c}\neq \mathbf{0}_{T^c}} \chi^{\Phi}_{\p{\mathbf{x}_T\circ \mathbf{z}_{T^c}, \mathbf{x}_T\circ \mathbf{z}_{T^c}}} }\cdot \p{\sum_{\mathbf{z}_{T^c}\in \ZfTc\atop \mathbf{z}_{T^c}\neq \mathbf{0}_{T^c}} \chi^{\Phi}_{\p{\mathbf{y}_T\circ \mathbf{z}_{T^c}, \mathbf{y}_T\circ \mathbf{z}_{T^c}}} } \\
      & = \frac12\p{\sum_{\mathbf{x}_T\in \ZfT} \sum_{\mathbf{z}_{T^c}\in \ZfTc \atop \mathbf{z}_{T^c}\neq \mathbf{0}_{T^c}}\chi^{\Phi}_{\p{\mathbf{x}_T\circ \mathbf{z}_{T^c}, \mathbf{x}_T\circ \mathbf{z}_{T^c}}} }^2 = \frac12\p{\sum_{\mathbf{x}\in \Zfn \atop \exists i\in T^c, x_i \neq 0} \chi^{\Phi}_{\mathbf{x}\mathbf{x}}}^2 = \left(\frac{\sqrt{2}}{2}\inf_{T^c}[\Phi]\right)^2.
    \end{align*}
    To summarize,
    \begin{align*}
      D\left(\Phi, \Phi_T\otimes \mathcal{I}_{T^c}\right) &= \f{1}{\sqrt{2}d}\norm{J(\Phi)-J(\Phi_T\otimes \mathcal{I}_{T^c})}_F
      \le \sqrt{\inf_{T^c}[\Phi]} + \f{\sqrt{2}}{2}\inf_{T^c}[\Phi].
    \end{align*} 
\end{proof}
\par
Theorem \ref{theo_main} utilizes the influence on the complement $T^c$, $\Inf_{T^c}[\Phi]$, to upper bound the distance between $\Phi$ and the $T$-junta process $\Phi_{T}\otimes\mathcal{I}_{T^c}$, which we regard as the $T$-junta approximation of $\Phi$. 
Therefore, we refer to $D(\Phi,\Phi_{T}\otimes\mathcal{I}_{T^c})$ as the junta-approximation error.
Given $\Inf_{T^c}[\Phi] = \delta$, this error satisfies $D(\Phi,\Phi_{T}\otimes\mathcal{I}_{T^c})\leq \epsilon$, where $\epsilon = \sqrt{\delta}+\delta/\sqrt{2}$. 
Solving for $\delta$ as a function of $\epsilon$ yields 
\begin{equation}
    \delta=f(\epsilon)=\left(\sqrt{\sqrt{2}\epsilon+\frac{1}{2}}-\frac{\sqrt{2}}{2}\right)^2.
    \label{eq_fe}
\end{equation}
Therefore, $\Inf_{T^c}[\Phi]\leq f(\epsilon)$ implies $D(\Phi,\Phi_{T}\otimes\mathcal{I}_{T^c})\leq \epsilon$. 
For $\epsilon=1$, we obtain $\delta=f(\epsilon)\approx 0.4576$. Thus, only $\Inf_{T^c}[\Phi]\leq 0.4576$ provides a non-trivial bound the junta-approximation error.
For a small $\epsilon$ ($0\leq\epsilon\ll 1 $), we expand $f(\epsilon)$ as follows
\begin{equation}
    f(\epsilon)= \epsilon^2-\sqrt{2}\epsilon^3+\frac{5}{2}\epsilon^4+\mathcal{O}(\epsilon^5),
    \label{eq_f_expand}
\end{equation}
such that $f(\epsilon)>\epsilon^2-\sqrt{2}\epsilon^3$ for small $\epsilon$. 
\par
We also give out a reachable lower bound on the junta-approximation error via influence as the following proposition.
\begin{proposition}
        Let $\Phi$ be a quantum process over $n$ qubits and let $T\subseteq [n]$ with complement $T^c$.
        Define the reduced subprocess on $T$ by $\Phi_T(\rho_T):=\Tr_{T^c}\lrb{\Phi\p{\rho_T\otimes \f{I_{T^c}}{2^{\abs{T^c}}}}}$. Then, $D\left(\Phi, \Phi_T\otimes\mathcal{I}_{T^c}\right)\geq \frac{\inf_{T^c}[\Phi]}{\sqrt{|A|}}\geq \frac{\inf_{T^c}[\Phi]}{\sqrt{4^{|T|}(4^{|T^c|}-1)}} $.
        Here, $A=\{ \mathbf{x}\in \Zfn,\ \mathbf{x}_{T^c}\neq \mathbf{0}_{T^c}: \chi^{\Phi}_{\mathbf{x}\mathbf{x}}\neq 0 \}$ and $\chi^{\Phi}$ denotes the process ($\chi$-)matrix of $\Phi$.
\end{proposition}
\begin{proof}
The $\chi$-matrix representation of $T$-junta $\Phi_{T}\otimes\mathcal{I}_{T^c}$ is given by
\begin{equation}
    \begin{split}
         \Phi_{T}\otimes\mathcal{I}_{T^c}(\rho) & =\sum_{\mathbf{x}_T,\mathbf{y}_T\in \ZfT} \chi^{\Phi_{T}\otimes\mathcal{I}_{T^c}}_{\p{\mathbf{x}_T\circ \mathbf{0}_{T^c}, \mathbf{x}_T\circ \mathbf{0}_{T^c}}} \left(\sigma_{\mathbf{x}_T}\otimes I_{\mathbf{x}_T^c}\right)\rho \left(\sigma_{\mathbf{y}_T}\otimes I_{\mathbf{x}_T^c}\right)  \\ 
         & =\sum_{\mathbf{x}_T,\mathbf{y}_T\in \ZfT} 
          \left(\sum_{\mathbf{z}_{T^c}\in \ZfTc}\chi^{\Phi}_{\p{\mathbf{x}_T\circ \mathbf{z}_{T^c}, \mathbf{x}_T\circ \mathbf{z}_{T^c}}}\right)
         \left(\sigma_{\mathbf{x}_T}\otimes I_{\mathbf{x}_T^c}\right)\rho \left(\sigma_{\mathbf{y}_T}\otimes I_{\mathbf{x}_T^c}\right).
    \end{split}
\end{equation}
Thus, the squared Frobenius norm of $ \chi^\Phi-\chi^{\Phi_T\otimes \mathcal{I}_{T^c}} $ can be written as
\begin{equation}
    \begin{split}
        &\quad\   \norm{\chi^\Phi-\chi^{\Phi_T\otimes \mathcal{I}_{T^c}}}_F^2 \\
        & = \sum_{\mathbf{x},\mathbf{y}\in \Zfn \atop \mathbf{x}_{T^c}\neq \mathbf{0}_{T^c}\ \mathrm{or}\ \mathbf{y}_{T^c}\neq \mathbf{0}_{T^c}} \left| \chi^{\Phi}_{\mathbf{x}\mathbf{y}}  \right|^2 + \sum_{\mathbf{x}_T,\mathbf{y}_T\in \ZfT}\left|\chi^{\Phi}_{\p{\mathbf{x}_T\circ \mathbf{0}_{T^c}, \mathbf{y}_T\circ \mathbf{0}_{T^c}}} - \sum_{\mathbf{z}_{T^c}\in \ZfTc}\chi^{\Phi}_{\p{\mathbf{x}_T\circ \mathbf{z}_{T^c}, \mathbf{y}_T\circ \mathbf{z}_{T^c}}}  \right|^2   \\
        & = \sum_{\mathbf{x},\mathbf{y}\in \Zfn \atop \mathbf{x}_{T^c}\neq \mathbf{0}_{T^c}\ \mathrm{or}\ \mathbf{y}_{T^c}\neq \mathbf{0}_{T^c}} \left| \chi^{\Phi}_{\mathbf{x}\mathbf{y}}  \right|^2 + \sum_{\mathbf{x}_T,\mathbf{y}_T\in \ZfT}\left|\sum_{\mathbf{z}_{T^c}\in \ZfTc \atop \mathbf{z}_{T^c}\neq \mathbf{0}_{T^c}}\chi^{\Phi}_{\p{\mathbf{x}_T\circ \mathbf{z}_{T^c}, \mathbf{y}_T\circ \mathbf{z}_{T^c}}}  \right|^2  \\
        & \geq  \sum_{\mathbf{x}\in \Zfn \atop \mathbf{x}_{T^c}\neq \mathbf{0}_{T^c}} \left| \chi^{\Phi}_{\mathbf{x}\mathbf{x}}  \right|^2 + \sum_{\mathbf{x}_T\in \ZfT}\left|\sum_{\mathbf{x}_{T^c}\in \ZfTc \atop \mathbf{x}_{T^c}\neq \mathbf{0}_{T^c}}\chi^{\Phi}_{\p{\mathbf{x}_T\circ \mathbf{x}_{T^c}, \mathbf{x}_T\circ \mathbf{x}_{T^c}}}  \right|^2 \quad   \p{\text{Removing off-diagonal terms} } \\
        & \geq  2\sum_{\mathbf{x}\in \Zfn \atop \mathbf{x}_{T^c}\neq \mathbf{0}_{T^c}} \left| \chi^{\Phi}_{\mathbf{x}\mathbf{x}}  \right|^2 \quad   \p{\chi^{\Phi}_{\mathbf{x}\mathbf{x}}\text{ is real and non-negative}}\\
        & \geq \frac{2}{|A|}\p{\sum_{\mathbf{x}\in A}\chi^{\Phi}_{\mathbf{x}\mathbf{x}}}^2 \quad \p{\text{Cauchy–Schwarz inequality}}.
    \end{split}
\end{equation}
Here, $A=\{ \mathbf{x}\in \Zfn,\ \mathbf{x}_{T^c}\neq \mathbf{0}_{T^c}: \chi^{\Phi}_{\mathbf{x}\mathbf{x}}\neq 0 \}$. 
Clearly, $|A|\leq 4^{|T|}(4^{|T^c|}-1)$. 
Therefore, the distance between $\Phi$ and $\Phi_T\otimes \mathcal{I}_{T^c}$ is bounded in terms of the influence by
\begin{equation}
    D\left(\Phi, \Phi_T\otimes \mathcal{I}_{T^c}\right)\geq \frac{\inf_{T^c}[\Phi]}{\sqrt{|A|}}\geq \frac{\inf_{T^c}[\Phi]}{\sqrt{4^{|T|}(4^{|T^c|}-1)}}.
\end{equation}
The first inequality becomes an equality if and only if the following hold simultaneously:
1. $\chi^\Phi$ is diagonal; 
2. For each fixed $\mathbf{x}_{T}$, at most one $\mathbf{x}_{T^c}\neq \mathbf{0}_{T^c}$ yields a nonzero $\chi_{\p{\mathbf{x}_T\circ \mathbf{x}_{T^c}, \mathbf{x}_T\circ \mathbf{x}_{T^c}}}$;
3. $\chi_{\mathbf{x}\mathbf{x}}^{\Phi} = \lambda$ for all $\mathbf{x}\in A$.
\end{proof}
\par
Consider a process $\Psi$ whose $\chi$-matrix saturates the first inequality.
Let its only nonzero diagonal entries be $\chi^{\Psi}_{\p{\mathbf{x}_T\circ \mathbf{z}_{T^c}, \mathbf{x}_T\circ \mathbf{z}_{T^c}}} = \lambda/4^{|T|}$, for all $\mathbf{x}_T\in \ZfT$, for a fixed $\mathbf{z}_{T^c}\neq \mathbf{0}_{T^c}$ and with $0\leq\lambda\leq 1$.
Then the influence on $T^c$ is $\inf_{T^c}[\Psi] = \lambda$.
For the $T$-junta $\Psi_T\otimes \mathcal{I}_{T^c}$, 
the nonzero diagonal entries are $\chi^{\Psi_T\otimes \mathcal{I}_{T^c}}_{\p{\mathbf{x}_T\circ \mathbf{0}_{T^c}, \mathbf{x}_T\circ \mathbf{0}_{T^c}}} = \lambda/4^{|T|}$, for all $\mathbf{x}_T\in \ZfT$. So $A$ has size $|A|=4^{|T|}$. 
The distance is therefore $D(\Psi,\Psi_T\otimes \mathcal{I}_{T^c})= \lambda/2^{|T|}=\inf_{T^c}[\Psi]/2^{|T|}$. 
This special case shows that a large influence does not, in general, imply a large junta-approximation distance for arbitrary quantum processes.
However, when the process is a unitary gate $\mathcal{U}$, the influence does yield a dimension-independent lower bound on the distance, allowing an efficient scheme for tolerant junta testing~\cite{bao2025efficient}.

\section{Influence Samplers and Influence Bounds}
\label{app_c}
In this section, we deduce the bounds on influence produced by influence samplers.
Given a quantum process $\Phi$ and a qubit set $S$, 
the influence sampler associated with a specific test gate $U_l$ is given by:
\begin{equation}
\begin{split}
\mathbb{E}X^{S}_l & = \mathop{\mathbb{E}}_{\T_l\sim p(\T_l)} \left\{\id \lrb{\T_l\cap S \neq \varnothing}\right\} \\
& = 1 - \frac{1}{2^n}\sum_{a\in\{0,1\}^{S}}\sum_{b\in\{0,1\}^{S^c}}\text{Tr}\p{\ketbra{a}_{S}\mathrm{Tr}_{S^c}\left[U_l^{\dagger\otimes n}\Phi\p{U_l^{\otimes n}\ketbra{a}_{S}\otimes\ketbra{b}_{S^c}U_l^{\dagger\otimes n}}U_l^{\otimes n}\right]}\\
& = 1 - \frac{1}{2^{|S|}}\sum_{a\in\{0,1\}^{S}}\text{Tr}\p{\ketbra{a}_{S}\mathrm{Tr}_{S^c}\left[U_l^{\dagger\otimes n}\Phi\p{U_l^{\otimes n}\ketbra{a}_{S}\otimes \frac{I_{S^c}}{2^{|S^c|}} U_l^{\dagger\otimes n}}U_l^{\otimes n}\right]}\\
\end{split}
\label{eq_S_ex}
\end{equation}
where $\id[\cdot]$ denotes the indicator function, $\ket{a}_S$ and $\ket{b}_{S^c}$ denote the computational bases over $S$ and its complement $S^c$, respectively. 
$\mathbb{E}X^{S}_l$ can be estimated from the empirical distribution $\hat{p}(\T_l)$ obtained by influence sampling.
Averaging over $p(\T_l)$ in equation~(\ref{eq_S_ex}) is equivalent to averaging the measurement probability over initial states where each qubit is independently and uniformly chosen from $\{\ket{0}, \ket{1}\}$. 
The simplification to using the maximally mixed state $I_{S^c}/2^{|S^c|}$ on $S^c$ in equation~(\ref{eq_S_ex}) arises because averaging over uniformly random computational basis states on $S^c$ after tracing out the qubits in $S^c$ is equivalent to using $I_{S^c}/2^{|S^c|}$ as input.
\par
The relationship between influence samplers and the process matrix can be derived as~\cite{bao2023testing}:
\begin{equation}
\begin{split}
    \mathbb{E}X^S_1 = 1 - \sum_{\substack{\forall i\in S,\  x_i \in \{0,3\} } }\chi^{\Phi}_{\mathbf{x}\mathbf{x}};\quad \mathbb{E}X^S_2 = 1 - \sum_{\substack{\forall i\in S,\  x_i \in \{0,1\} } }\chi^{\Phi}_{\mathbf{x}\mathbf{x}};\quad  \mathbb{E}X^S_3 = 1 - \sum_{\substack{\forall i\in S,\  x_i \in \{0,2\} } }\chi^{\Phi}_{\mathbf{x}\mathbf{x}}.
\end{split}
\end{equation}
For convenience, we introduce the following notations for various partial sums of the diagonal elements of $\chi^{\Phi}$: 
\begin{equation}
\begin{split}
    O & = \sum_{\substack{\forall i\in S,\  x_i=0 } }\chi^{\Phi}_{\mathbf{x}\mathbf{x}} = 1-\Inf_S[\Phi],\quad A = \sum_{\substack{\forall i\in S,\  x_i \in \{0,3\} } }\chi^{\Phi}_{\mathbf{x}\mathbf{x}} = 1 - \mathbb{E}X^S_1,\\ 
    B & = \sum_{\substack{\forall i\in S,\  x_i \in \{0,1\} } }\chi^{\Phi}_{\mathbf{x}\mathbf{x}}=1 - \mathbb{E}X^S_2,\quad C = \sum_{\substack{\forall i\in S,\  x_i \in \{0,2\} } }\chi^{\Phi}_{\mathbf{x}\mathbf{x}} = 1 - \mathbb{E}X^S_3,
\end{split}
\end{equation}
and their corresponding difference terms:
\begin{equation}
    \begin{split}
         A_o & = A-O = \sum_{\substack{\forall i\in S,\  x_i \in \{0,3\}; \\ \exists j\in S, x_j \neq 0 } }\chi^{\Phi}_{\mathbf{x}\mathbf{x}},\  B_o = B-O = \sum_{\substack{\forall i\in S,\  x_i \in \{0,1\}; \\ \exists j\in S, x_j \neq 0 } }\chi^{\Phi}_{\mathbf{x}\mathbf{x}},\ 
         C_o = C-O = \sum_{\substack{\forall i\in S,\  x_i \in \{0,2\}; \\ \exists j\in S, x_j \neq 0 } }\chi^{\Phi}_{\mathbf{x}\mathbf{x}},\\
        D & = 1-(O+A_o+B_o+C_o) = 1-(A+B+C-2O) = \sum_{\substack{\exists i,j\in S,\ x_i, x_j\in \{1,2,3\},\\ x_i\neq x_j } }\chi^{\Phi}_{\mathbf{x}\mathbf{x}}.
    \end{split}
\end{equation}
By construction, all of $O,A,B,C,A_o,B_o,C_o,D$ are non-negative.
The identity is expressed as $O+A_o+B_o+C_o+D=1$.
Firstly, we show that any two of the three influence samplers yield valid bounds on the influence.
Considering $\left\{\mathbb{E}X^S_1,\mathbb{E}X^S_2\right\}$ as an example, the influence inequality derived from two samplers is:
\begin{equation}
\begin{split}
    & \frac{A+B}{2} \geq \mathop{\min}\{A,B\} \geq O\geq O-(C_o+D) = A+B-1 \\
    \Rightarrow &\  \frac{1-A+1-B}{2} \leq \mathop{\max}\{1-A,1-B\}  \leq 1-O \leq 1-A + 1-B \\
    \Rightarrow &\ \frac{1}{2}\left(\mathbb{E}X^S_1+\mathbb{E}X^S_2\right)  \leq \mathop{\max}\{\mathbb{E}X^S_1,\mathbb{E}X^S_2\} \leq \text{Inf}_{S}[\Phi] \leq \mathbb{E}X^S_1+\mathbb{E}X^S_2.
\end{split}
\end{equation}
Analogous inequalities hold for $\left\{\mathbb{E}X^S_2,\mathbb{E}X^S_3\right\}$ and $\left\{\mathbb{E}X^S_1,\mathbb{E}X^S_3\right\}$.
\par
The influence inequality derived from three influence samplers is:
\begin{equation}
\begin{split}
     & \frac{A+B+C}{3} \geq \mathop{\min}\{A,B,C\}  \geq O\geq O-\frac{D}{2}= \frac{A+B+C-1}{2} \\
    \Rightarrow &\ \frac{1-A+1-B+1-C}{3}\leq \mathop{\max}\{1-A,1-B,1-C\} \leq 1-O \leq \frac{1}{2}(1-A + 1-B + 1-C)\\
    \Rightarrow &\ \frac{1}{3}\left(\mathbb{E}X^S_1+\mathbb{E}X^S_2+ \mathbb{E}X^S_3\right)\leq \mathop{\max}\{\mathbb{E}X^S_1,\mathbb{E}X^S_2,\mathbb{E}X^S_3\} \leq \text{Inf}_{S}[\Phi] \leq \frac{1}{2}\left(\mathbb{E}X^S_1+\mathbb{E}X^S_2+ \mathbb{E}X^S_3\right).
\end{split}
\end{equation}
We denote the influence bounds obtained from these samplers as
\begin{equation}
\begin{split}
     \mathrm{IL}_S & = \mathop{\max}\{\mathbb{E}X^S_1,\mathbb{E}X^S_2\},\quad \mathrm{IU}_S = \mathbb{E}X^S_1+\mathbb{E}X^S_2;\\ 
     \text{IL}_S^{\mathrm{(II)}} & = \mathop{\max}\{\mathbb{E}X^S_1,\mathbb{E}X^S_2,\mathbb{E}X^S_3\},\quad \text{IU}_S^{\mathrm{(II)}} = \frac{1}{2}\left(\mathbb{E}X^S_1+\mathbb{E}X^S_2+ \mathbb{E}X^S_3\right).
\end{split}
\label{eq_ib_all}
\end{equation}
Obviously, $\mathrm{IL}_S\leq \text{IL}_S^{\mathrm{(II)}}$.
From $D/2\leq C_o + D$, we obtain $\mathrm{IU}_S^{\mathrm{(II)}}\leq \text{IU}_S$. 
Hence, incorporating $\mathbb{E}X^S_3$ yields tighter upper and lower bounds on influence.
In summary, the influence inequalities can be written as
\begin{equation}
\begin{split}
\frac{1}{2}\text{IU}_S \leq \text{IL}_S \leq \text{Inf}_{S}[\Phi] \leq \text{IU}_S;\quad &
\frac{2}{3}\text{IU}_S^{\mathrm{(II)}}  \leq \text{IL}_S^{\mathrm{(II)}} \leq \text{Inf}_{S}[\Phi] \leq \text{IU}_S^{\mathrm{(II)}};\\
\text{IL}_S\leq \text{IL}_S^{\mathrm{(II)}} \leq \text{Inf}_{S}[\Phi] & \leq \text{IU}_S^{\mathrm{(II)}} \leq \text{IU}_S.
\end{split}
\label{ieq_inf}
\end{equation}
This completes the proof of equations~(\ref{ieq_1}) and (\ref{ieq_2}).
\par
Furthermore, we can use the influence true value, $\operatorname{Inf}_S[\Phi]$, to bound $\mathrm{IU}_S$ and $\mathrm{IU}_S^{\mathrm{(II)}}$:
\begin{equation}
\begin{split}
    \inf_S[\Phi] & \leq \text{IU}_S \leq 2\inf_S[\Phi],\\
    \inf_S[\Phi] & \leq \text{IU}_S^{\mathrm{(II)}} \leq \frac{3}{2}\inf_S[\Phi].
\end{split}
\end{equation}
These show that influence upper bounds provide valid approximation of the true influence within constant factors.

\section{Sample complexity of influence sampling}
\label{app_d}
In this section, we analyze the sample complexity of influence sampling, from the scenario that using $M$ samples per test gate to estimate the influence upper bounds on $T^c$, thereby obtaining a bound on the junta‑approximation distance $D(\Phi,\Phi_T\otimes\mathcal{I}_{T^c})\leq \epsilon$.
For brevity, we write $\mathrm{IU}$, $\mathrm{IU}^{\mathrm{(II)}}$ and $\Inf$ to denote the true values $\mathrm{IU}_{T^c}$, $\mathrm{IU}^{\mathrm{(II)}}_{T^c}$ and $\Inf_{T^c}[\Phi]$.
Given raw influence samplers $\{\{X_l[j]\}_{j=1}^{M}\}_{l=1}^{3}$,
define $Y_1[j] := X_1[j]+X_2[j]$, $Y_2[j] := (X_1[j]+X_2[j]+X_3[j])/2$, so that $\mathbb{E}Y_1 = \mathrm{IU}$, $Y_1\in \{0,1,2\}$; $\mathbb{E}Y_2 = \mathrm{IU}^{\mathrm{(II)}}$, $Y_2\in \left\{0,\frac{1}{2},1,\frac{3}{2}\right\}$. 
Let $\overline{Y}_i:=\sum_jY_i[j]/M$ denote the empirical influence upper bounds.
\par
Given a target distance bound $\epsilon$, the corresponding influence threshold is $\delta=f(\epsilon)$ from equation~\ref{eq_fe}.
By theorem~\ref{theo_main}, $\Inf\leq\delta$ implies $D(\Phi,\Phi_T\otimes\mathcal{I}_{T^c})\leq \epsilon$.
Choose a decision threshold $\delta_0\leq \delta$ and accept only if $\overline{Y}_i\leq\delta_0$. 
A false acceptance occurs when $\Inf\geq \delta$ but $\overline{Y}_i\leq\delta_0$. 
We require the probability of any such error to be at most $\eta$. 
Consequently, if we observe $\overline{Y}_i\leq\delta_0$, we conclude that $D(\Phi,\Phi_T\otimes\mathcal{I}_{T^c})\leq \epsilon$ with confidence level at least $1-\eta$.
\par
\textbf{Sample complexity via Hoeffiding inequality.---} 
Applying Hoeffding’s inequality in each $\overline{Y}_i$ gives
\begin{equation}
\adjustbox{scale=0.9}{$
\begin{split}
    \Pr(\overline{Y}_1\leq\delta_0,\ \Inf> \delta) & \leq \Pr( \overline{Y}_1\leq\delta_0,\ \mathrm{IU}> \delta)\leq \Pr(\mathrm{IU}-\overline{Y}_1\geq \delta-\delta_0)\leq \exp(-\frac{1}{2}M(\delta-\delta_0)^2)\leq \eta, \\
    \Pr(\overline{Y}_2\leq\delta_0,\ \Inf> \delta) & \leq \Pr(\overline{Y}_2\leq\delta_0,\ \mathrm{IU}^{(\mathrm{II})}> \delta)\leq \Pr(\mathrm{IU}^{(\mathrm{II})}-\overline{Y}_2\geq \delta-\delta_0)\leq \exp(-\frac{8}{9}M^{(\mathrm{II})}(\delta-\delta_0)^2)\leq \eta. \\
\end{split}
$}
\end{equation}
To drive each of these probabilities below $\eta$, it suffices to choose
\begin{equation}
    M\geq \frac{2\ln(1/\eta)}{(\delta-\delta_0)^2},\quad M^{(\mathrm{II})}\geq\frac{9\ln(1/\eta)}{8(\delta-\delta_0)^2}.
\end{equation}
For $\epsilon=0.1$, $\delta=f(\epsilon)\approx 8.79\times10^{-3}$, $\eta=1/3$, $\delta_0 = 0.7\delta$, we have $M\geq 3.16\times 10^5$ and $M^{(\mathrm{II})}\geq 1.78\times 10^5$. 
\par
\textbf{Sample complexity via large deviation analysis.---}Since $\delta$ is also small for a small $\epsilon$, we employ a large deviation principle with worst-case two-point reduction to obtain sharper tail bounds and hence tighter sample complexity estimates.
Firstly, we observe that
\begin{equation}
    \Pr(\overline{Y}_1\leq\delta_0,\ \Inf> \delta)\leq \Pr(\overline{Y}_1\leq\delta_0,\ \mathrm{IU}> \delta) \leq 
    \sup_{P:\ \mathrm{IU}> \delta} \Pr_P(\overline{Y}_1\leq\delta_0),
\end{equation}
and similarly $\Pr(\overline{Y}_2\leq\delta_0,\ \Inf> \delta)\leq\sup_{P:\ \mathrm{IU}^{(\mathrm{II})}> \delta} \Pr_P(\overline{Y}_2\leq\delta_0)$.
The worst-case distributions are supported on the endpoints $\{0,2\}$ for $Y_1$ and $\left\{0,\frac{3}{2}\right\}$ for $Y_2$. 
Large deviation principle yields the following inequalities~\cite{Dembo2010}:
\begin{equation}
    \begin{split}
        \sup_{P:\ \mathrm{IU}> \delta} \Pr_P(\overline{Y}_1\leq\delta_0) & \leq \exp[- D_{\mathrm{KL}}\left(\frac{\delta_0}{2}\Big\Vert \frac{\delta}{2} \right)M], \\
         \sup_{P:\ \mathrm{IU}^{(\mathrm{II})}> \delta} \Pr_P(\overline{Y}_2\leq\delta_0) & \leq \exp[- D_{\mathrm{KL}}\left(\frac{2\delta_0}{3}\Big\Vert \frac{2\delta}{3} \right)M^{(\mathrm{II})}], 
    \end{split}
\end{equation}
where $D_{\mathrm{KL}}(p\Vert q)=p\ln(p/q)+(1-p)\ln[(1-p)/(1-q)]$ is
the Kullback-Leibler divergence.
 Requiring each error $\leq \eta$ gives
\begin{equation}
   M\geq \frac{\ln(1/\eta)}{D_{\mathrm{KL}}\left(\frac{\delta_0}{2}\Big\Vert \frac{\delta}{2} \right)},\quad M^{(\mathrm{II})}\geq \frac{\ln(1/\eta)}{D_{\mathrm{KL}}\left(\frac{2\delta_0}{3}\Big\Vert \frac{2\delta}{3} \right)}.
\end{equation}
For $\delta\approx 8.79\times10^{-3}$, $\delta_0 = 0.7\delta$, and $\eta=1/3$, this yields numerically $M\geq 4.95\times 10^3$ and $M^{(\mathrm{II})}\geq 3.71\times 10^3$.
Since each sample of $Y_1$ requires two queries and each sample of $Y_2$ requires three, the total query counts satisfy
\begin{equation}
    M_{t}=2M\geq\frac{2\ln(1/\eta)}{D_{\mathrm{KL}}\left(\frac{\delta_0}{2}\Big\Vert \frac{\delta}{2} \right)},\quad M_{t}^{(\mathrm{II})} = 3M^{(\mathrm{II})}\geq \frac{3\ln(1/\eta)}{D_{\mathrm{KL}}\left(\frac{2\delta_0}{3}\Big\Vert \frac{2\delta}{3} \right)}.
\end{equation}
\par
For small $p$, $q$ and $p\lesssim q$,  we can use the quadratic approximation $D_{\mathrm{KL}}(p\Vert q)\approx(p-q)^2/2q$. Let $\delta-\delta_0 = \delta/\kappa$, we have
\begin{equation}
    M\geq  \frac{4\delta\ln(1/\eta)}{(\delta-\delta_0)^2} = \frac{4\kappa^2\ln(1/\eta)}{\delta},\quad M^{(\mathrm{II})}\geq \frac{3\delta\ln(1/\eta)}{(\delta-\delta_0)^2}=\frac{3\kappa^2\ln(1/\eta)}{\delta}.
\end{equation}
Therefore the sample complexity for influence sampling scales as $M=\mathcal{O}(\kappa^2\ln(1/\eta)/\delta)$ with $0<\delta_0\lesssim\delta\ll 1$.
From equation~\ref{eq_f_expand}, since $\delta=f(\epsilon)=\mathcal{O}(\epsilon^2)$ for small $\epsilon$, 
we further obtain $M=\mathcal{O}(\kappa^2\ln(1/\eta)/\epsilon^2)$.
\par

\section{Algorithms of influence sampling}
\label{app_e}
In this section, we present the influence sampling in the algorithmic form.
We first introduce $\texttt{Influence-Sample}$ for a fixed test gate $U_l$, producing a dataset $\mathcal{K}_l$ consisting of $M$ samples $\{\mathcal{T}_l^{(j)}\}_{j=1}^{M}$, as shown in Algorithm~\ref{alg:IS}. 
In the quantum sampling stage, we invoke \texttt{Influence-Sample} for different $U_l$ to get the collection of datasets $\{\mathcal{K}_l\}$.
\par
\begin{algorithm}[h]
\caption{$\texttt{Influence-Sample}(\Phi, M, l)$}
\label{alg:IS}
\SetKwInOut{Input}{Input}
\SetKwInOut{Output}{Output}
\Input{Oracle access to quantum process $\Phi$, test gate index $l\in\{1,2,3\}$, and repetitions $M$.}
\Output{A dataset of $M$ samples of high-influence qubit subsets for $U_l$: $\mathcal{K}_l = \{\mathcal{T}^{(j)}_l\}_{j=1}^M$, with $\mathcal{T}^{(j)}_l\subseteq [n]$.}
1. Select the test gate $U_l$ according to $l$ from
\begin{equation*}
\begin{aligned}[b]
        \left\{ U_1: I = \begin{pmatrix} 1 & 0 \\ 0 & 1 \end{pmatrix},\quad 
        U_2: H = \frac{1}{\sqrt{2}} \begin{pmatrix} 1 & 1 \\ 1 & -1 \end{pmatrix},\quad
        U_3: R_x\left(\frac{\pi}{2}\right) = \frac{1}{\sqrt{2}} \begin{pmatrix} 1 & -i \\ -i & 1 \end{pmatrix} \right\};
\end{aligned}
\end{equation*}
2. Initialize $\mathcal K_l\!\leftarrow\!\varnothing$\;  
\For{$j=1$ \KwTo\ $M$}{
    3. Uniformly randomly choose $\displaystyle a \in \{0, 1\}^n$ and prepare the state $\displaystyle \ket{a}$\;
    4. Obtain $\displaystyle \left(U_l^{\otimes n}\right)^\dagger \Phi \left( U_l^{\otimes n} \ket{a}\!\!\bra{a} \left(U_l^{\otimes n}\right)^\dagger \right) U_l^{\otimes n}$\;
    5. Measure all qubits in the computational basis and record the outcome as $\displaystyle b$\;  
    6. Set $\displaystyle \mathcal{T}^{(j)}_l = \{i \in [n] \mid a_i \neq b_i\}$\;  
    7. Update $\displaystyle \mathcal{K}_l \gets \mathcal{K}_l \cup \{\mathcal{T}_l^{(j)}\}$\;  
}
8. \Return $\mathcal K_l$  
\end{algorithm}
\par
Next, we provide introduce an alternative influence-sampling scheme  that randomly selects the test gates, as described in Algorithm~\ref{alg:IS-RU}.
By invoking \texttt{Influence-Sample-RU}, we obtain a dataset $\mathcal{K}_{\mathrm{I}}$ or $\mathcal{K}_{\mathrm{II}}$, corresponding to two different choices of random testing gates.
Given $\mathcal{K}_{\mathrm{I}}$ or $\mathcal{K}_{\mathrm{II}}$, we can estimate the probability of the sampled subsets overlapping with a given set $S$.
These probabilities can be expressed in terms of the influence samplers as
\begin{equation}
\begin{split}
    \Prob{\mathcal{T}_{\mathrm{I}}\cap S\neq \varnothing} = \frac{1}{2}\left(\mathbb{E}X^S_1+\mathbb{E}X^S_2\right),\ \ \Prob{\mathcal{T}_{\mathrm{II}}\cap S\neq \varnothing} = \frac{1}{3}\left(\mathbb{E}X^S_1+\mathbb{E}X^S_2 + \mathbb{E}X^S_3\right).
\end{split}
\end{equation}
With influence inequalities in equation~(\ref{ieq_inf}), we can rewrite the influence bounds in terms of these overlap probabilities:
\begin{equation}
\begin{split}
    \frac{1}{2}\mathrm{IU}_S=\Prob{S\cap \mathcal{T}_{\mathrm{I}}\neq \varnothing} & \leq \inf_S[\Phi] \leq 2\Prob{S\cap \mathcal{T}_{\mathrm{I}}\neq \varnothing} = \mathrm{IU}_S,\\
    \frac{2}{3}\mathrm{IU}_S^{\mathrm{(II)}} = \Prob{S\cap \mathcal{T}_{\mathrm{II}}\neq \varnothing} & \leq \inf_S[\Phi] \leq \frac{3}{2}\Prob{S\cap \mathcal{T}_{\mathrm{II}}\neq \varnothing}=\mathrm{IU}_S^{\mathrm{(II)}}.
\end{split}
\end{equation}
Comparing these bounds with equation~(\ref{eq_ib_all}), we observe that both algorithms yield the same upper bounds on the influence, namely $\mathrm{IU}_S$ and $\mathrm{IU}_S^{\mathrm{(II)}}$. 
However, the lower bounds $\mathrm{IL}_S$ and $\mathrm{IL}_S^{\mathrm{(II)}}$ in equation~(\ref{eq_ib_all}) are tighter, since they are defined via a maximization over all influence samplers, which cannot be exactly estimated using \texttt{Influence-Sample-RU}.
\begin{algorithm}[ht]
\SetKwInOut{Input}{Input}
\SetKwInOut{Output}{Output}
\caption{$\texttt{Influence-Sample-RU}(\Phi, M, R)$}
\label{alg:IS-RU}
\Input{Oracle access to quantum process $\Phi$, test gate set indicator $R\in\{\text{I}, \text{II}\}$, and repetitions $M$.}
\Output{A dataset of $M$ samples of high-influence qubit subsets: $\mathcal{K}_R = \{\mathcal{T}^{(j)}_R\}_{j=1}^M$, with $\mathcal{T}^{(j)}_R\subseteq [n]$.}
1. Initialize $\mathcal{K}_R\!\leftarrow\!\varnothing$\;
\For{$j=1$ \KwTo\ $M$}{
2. Uniformly randomly choose $a\in \{0,1\}^n$ and prepare the state $\ket{a}$\;
3. Uniformly randomly choose a test gate $U_l$ from the set:
    $\begin{cases}
    \{I, H\} & \text{if } R = \text{I} \\
    \{I, H, R_x\p{\frac{\pi}{2}}\} & \text{if } R = \text{II}
    \end{cases}$\;
4. Obtain $\displaystyle \left(U_l^{\otimes n}\right)^\dagger \Phi \left( U_l^{\otimes n} \ket{a}\!\!\bra{a} \left(U_l^{\otimes n}\right)^\dagger \right) U_l^{\otimes n}$\;
5. Measure all qubits in the computational basis and record the outcome as $\displaystyle b$\;  
6. Set $\displaystyle \mathcal{T}^{(j)}_R = \{i \in [n] \mid a_i \neq b_i\}$\;  
7. Update $\displaystyle \mathcal{K}_R \gets \mathcal{K}_l \cup \{\mathcal{T}_R^{(j)}\}$\;  
}
8. \Return $\mathcal{K}_R$  
\end{algorithm}
\par
We now describe how to identify a high-influence qubit subset $T$ using the datasets $\mathcal{K}_l$ produced by $\texttt{Influence-Sample}$, as summarized in Algorithm~\ref{alg:HIQI}.
Note that this algorithm no longer queries the quantum process, representing a classical post-processing procedure.
This procedure can also take $\{\mathcal{K}_1, \mathcal{K}_2, \mathcal{K}_3\}$ as input, in which case it leverages the tighter influence bounds $\mathrm{IU}^{\mathrm{(II)}}$.
The output $\mathrm{IU}_{T^c}$ of Algorithm~\ref{alg:HIQI} can then be used to bound the $T$-junta approximation error via theorem~\ref{theo_main}.
\begin{algorithm}[ht]
\caption{\texttt{High-Inf-Qubit-Identify}($\{\mathcal{K}_1, \mathcal{K}_2\}$, $\delta_0$)}
\SetKwInOut{Input}{Input}
\SetKwInOut{Output}{Output}
\label{alg:HIQI}
\Input{Collection of datasets $\{\mathcal{K}_1, \mathcal{K}_2\}$; decision threshold $\delta_0$ for single-qubit influence.}
\Output{High-influence qubit subset $T$ and an influence upper bound $\mathrm{IU}_{T^c}$ for its complement $T^c$.}
1. Set $S = \{i\}$ for each qubit $i \in [n]$, and compute the datasets of raw influence samplers $\{X_1^{\{i\}}[j]\}$ and $\{X_2^{\{i\}}[j]\}$ from $\mathcal{K}_1$ and $\mathcal{K}_2$, respectively\;
2. From $\{X_1^{\{i\}}[j]\}$ and $\{X_2^{\{i\}}[j]\}$, compute the empirical influence samplers $\mathbb{E}X_1^{\{i\}}$ and $\mathbb{E}X_2^{\{i\}}$ on each qubit\;
3. Compute the influence upper bound on each qubit:
        $\mathrm{IU}_{\{i\}} = \mathbb{E}X_1^{\{i\}} + \mathbb{E}X_2^{\{i\}}$\;
4. Identify the set of high-influence qubits:
        $T \leftarrow \{i \in [n] \mid \mathrm{IU}_{\{i\}} > \delta_0\}$\;
5. Set $S = T^c$, compute the datasets of raw influence samplers $\{X_1^{T^c}[j]\}$ and $\{X_2^{T^c}[j]\}$ from $\mathcal{K}_1$ and $\mathcal{K}_2$, respectively\;
6. From $\{X_1^{T^c}[j]\}$ and $\{X_2^{T^c}[j]\}$, compute the empirical influence samplers $\mathbb{E}X_1^{T^c}$ and $\mathbb{E}X_2^{T^c}$ on $T^c$\;
7. Compute the influence upper bound on $T^c$: $\mathrm{IU}_{T^c}=\mathbb{E}X_1^{T^c}+\mathbb{E}X_2^{T^c}$ \;
8. \Return $T$ and $\mathrm{IU}_{T^c}$.
\end{algorithm}

\section{Photon source}
\label{app_f}
Light pulses with 150 fs duration, centered at 830 nm, from a ultrafast Ti:sapphire Laser (Coherent Mira-HP; 76MHz repetition rate) are firstly frequency doubled in a $\beta$-type barium borate ($\beta$-BBO) crystal to generate a second harmonic beam with 415 nm wavelength. 
Then the upconversion beam is then utilized to pump another $\beta$-BBO with phase-matched cut angle for type-\uppercase\expandafter{\romannumeral2} beam-like degenerate spontaneous down conversion (SPDC) which produces pairs of photons, denoted as signal and idler. 
The signal and idler photons possess distinct emergence angles and spatially separate from each other. 
After passing through two clean-up filters with a 3nm bandwidth, the photons are coupled into separate single-mode fibers. 
The signal and idler photons are directed towards two separable linear optical circuits to under state preparation, evolution and final measurement by several single photon counting modules (SPCM, Excelitas Technologies) with a detection efficiency of approximately 60\%.
\section{State preparation and test gate implementation}
\label{app_g}
\begin{figure*}[ht]
    \centering
    \includegraphics[width= 0.9 \textwidth]{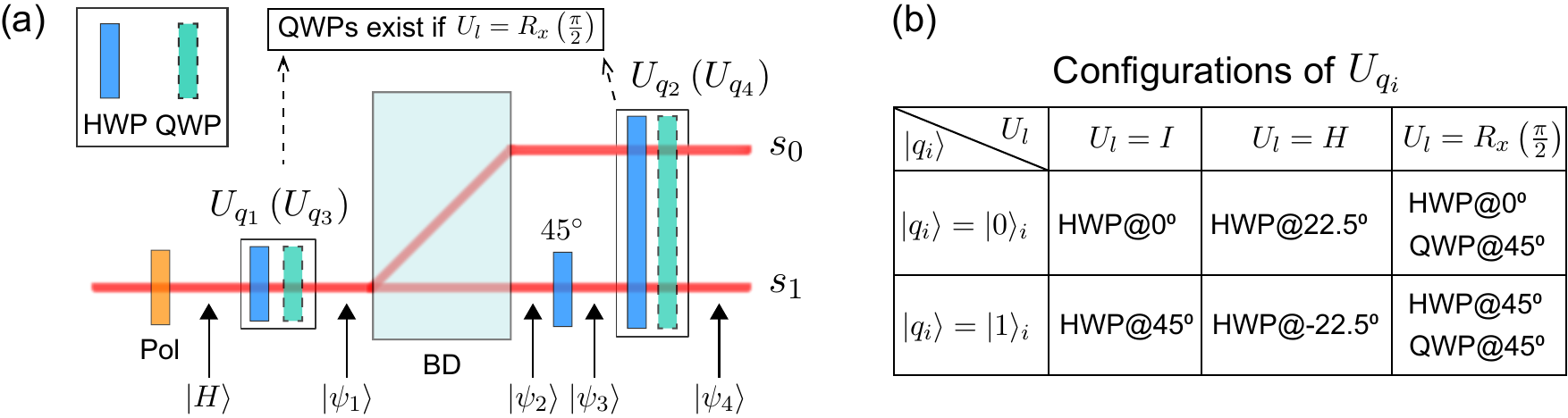}
    \caption{State preparation and test gate implementation for the first ($q_1$) and second ($q_2$)qubits. (a) Experimental setup. (b) Configurations of wave-plates for different initial states and test gates.
    Identical setups and encoding approaches are used for the third ($q_3$) and fourth ($q_4$) qubits.}
    \label{SM_fig4_SP}
\end{figure*}
This section details the experimental realization of state preparation and test gate implementation.
Our experimental setup consists of two separable two-qubit circuits, each receiving one photon from a photon pair as input. Within each two-qubit circuit, the input single photons are initialized to the horizontal polarization state $\ket{H}$ through a polarizer (Pol). Subsequently, state preparation and the test gate $U_l$ are firstly applied to the first qubit. This is achieved using a half-wave plate (HWP) for $U_l = I$ or $U_l = H$, and a combination of an HWP and a quarter-wave plate (QWP) for $U_l = R_x(-\pi/2)$. The HWP and QWP operations on the polarization are described by:
\begin{equation}
    U_{\text{HWP}}(\theta)=
    \p{\begin{matrix}
        \cos{2\theta} & \sin{2\theta} \\
        \sin{2\theta} & -\cos{2\theta}
    \end{matrix}},\quad
    U_{\text{QWP}}(\theta)=
    \p{\begin{matrix}
        \cos^2{\theta}+i\sin^2{\theta} & (1-i)\sin{\theta}\cos{\theta} \\
        (1-i)\sin{\theta}\cos{\theta} & \sin^2{\theta}+i\cos^2{\theta}
    \end{matrix}}.
\end{equation}
The transformation implemented by the wave plates on the first qubit ($q_1$) is denoted as $U_{q_1}$ in figure~\hyperref[SM_fig4_SP]{4(a)}, resulting in the single-photon state:
\begin{equation}
    \ket{\psi_1} = U_{q_1}\ket{H} = \alpha\ket{H}+\beta\ket{V},
\end{equation}
where $\alpha$ and $\beta$ are coefficients of a general polarization qubit with $|\alpha|^2+|\beta|^2=1$. The beam displacer (BD) separates the polarization qubit into the upper ($s_0$) and the lower ($s_1$) path modes, creating a polarization-path entangled state as
\begin{equation}
    \ket{\psi_2} = \alpha\ket{s_0}\otimes\ket{H}+\beta\alpha\ket{s_1}\otimes\ket{V}.
\end{equation}
The HWP at $45^{\circ}$ in path $s_1$ evolves the state to
\begin{equation}
    \ket{\psi_3} = \alpha\ket{s_0}\otimes\ket{H}+\beta\ket{s_1}\otimes\ket{H} = \p{\alpha\ket{s_0}+\beta\ket{s_1}}\otimes\ket{H}.
\end{equation}
Here, the path modes $\{\ket{s_0},\ket{s_1}\}$ represent the basis states of the first qubit ($q_1$), and the polarization modes $\{\ket{H},\ket{V}\}$ represent the basis states of the second qubit ($q_2$). 
The two-qubit basis states are encoded as
\begin{equation}
    \left\{\ket{0}_{1}\ket{0}_{2},\ket{0}_{1}\ket{1}_{2},\ket{1}_{1}\ket{0}_{2},\ket{1}_{1}\ket{1}_{2}\right\}\Rightarrow\left\{\ket{s_0}\otimes\ket{H},\ket{s_0}\otimes\ket{V},\ket{s_1}\otimes\ket{H},\ket{s_1}\otimes\ket{V}\right\}.
\end{equation}
Thus, $\ket{\psi_3}$ can be expressed as $\ket{\psi_3} = \p{U_{q_1}\ket{0}_{1}}\otimes\ket{0}_{2}$, where $U_{q_1}$ encode the fist qubit on the path modes and implements the test gate $U_l$. A similar unitary $U_{q_2}$, using the same wave plate configuration as $U_{q_1}$, is applied to the second qubit, resulting in the two-qubit state:
\begin{equation}
    \ket{\psi_4} = U_{q_1}\ket{0}_1\otimes U_{q_2}\ket{0}_2.
\end{equation}
The unitaries $U_{q_1}$ and $U_{q_2}$ combine state preparation and the test gate $U_l$ as following:
\begin{equation}
    \ket{\psi_4} =  U_{q_1}\ket{0}_1\otimes U_{q_2}\ket{0}_2 = U_l\ket{q_1}\otimes U_l\ket{q_2},\quad \ket{q_i} = \ket{0}_i\ \text{or}\ \ket{1}_i,\quad i = 1,2.
\end{equation}
Multiple configurations of $U_{q_i}$ ($i = 1,2$), as shown in figure~\hyperref[SM_fig4_SP]{4(b)}, satisfy $U_{q_i}\ket{0}_i = U_l\ket{q_i}$ for different prepared qubit states $\ket{q_i}$ and the test gate $U_l$. The identical setup and procedure are then applied to the remaining two qubits ($q_3$ and $q_4$).

\section{Construction of quantum processes}
\label{app_h}
\begin{figure*}[ht]
    \centering
    \includegraphics[width=0.9\textwidth]{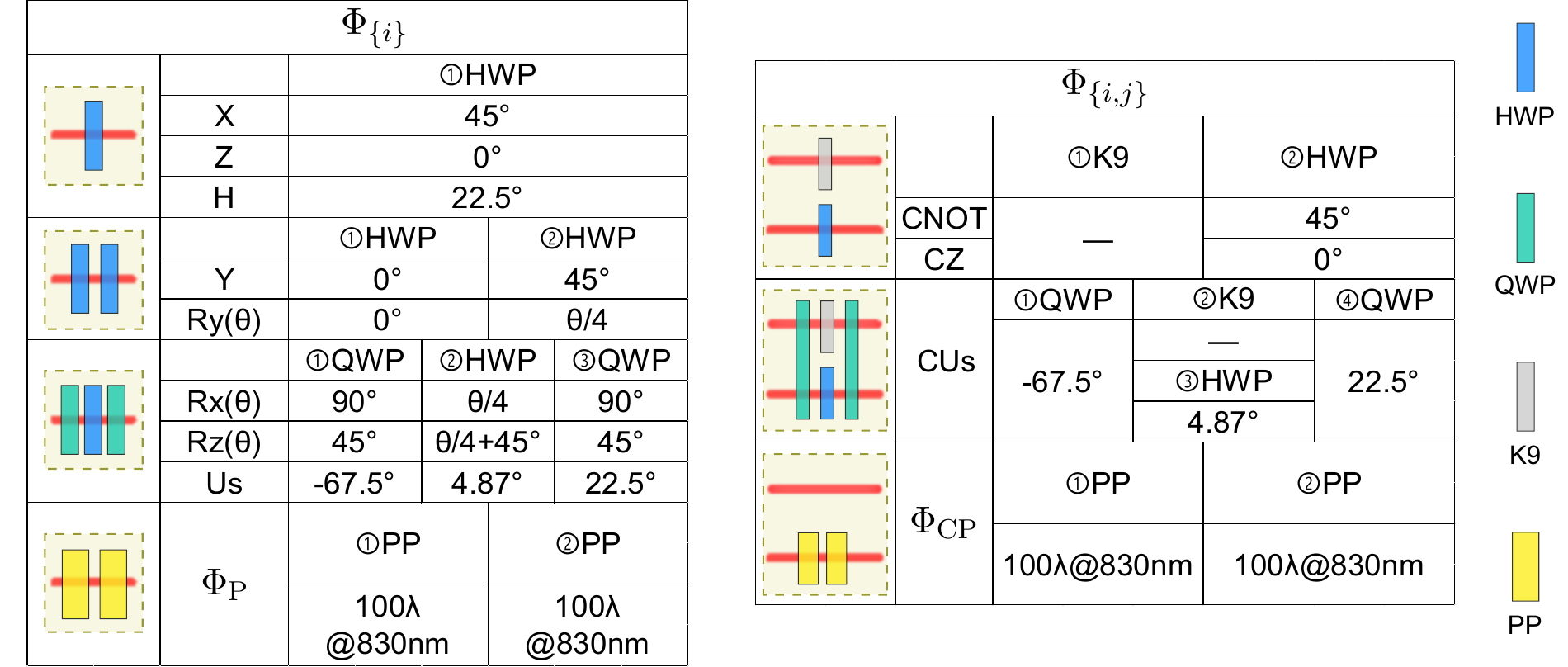}
    \caption{Experimental realizations of quantum subprocesses. The order of optical element labels is from left to right, from top to bottom. Here, the schematics of $\Phi_{\{i\}}$ also denote the same optical elements acting on both path modes. For two-qubit subprocesses $\Phi_{\{i,j\}}$, the path qubit is the control and the polarization qubit is the target. K9 glass compensates for optical path differences in CNOT, CZ and $\mathrm{CU_s}$. K9: K9-glass, PP: phase plate.}
    \label{SM_fig5_Channel}
\end{figure*}
This section details the optical implementation of the quantum processes used in our influence sampling experiments. The target processes, representing the unknown processes to be sampled for influence, are composed of fundamental single-qubit and two-qubit subprocesses. For single-qubit processes, we consider the four basic unitary gates $X$, $Y$, $Z$, and $H$ (Hadamard), along with three rotation gates:
\begin{equation}
    R_x(\theta) = \left( \begin{matrix} \cos{\frac{\theta}{2}} & -i\sin{\frac{\theta}{2}}\\ -i\sin{\frac{\theta}{2}} & \cos{\frac{\theta}{2}}\end{matrix}\right),\ R_y(\theta) = \left( \begin{matrix} \cos{\frac{\theta}{2}} & -\sin{\frac{\theta}{2}}\\ \sin{\frac{\theta}{2}} & \cos{\frac{\theta}{2}}\end{matrix}\right),\ 
    R_z(\theta) = \left( \begin{matrix} e^{-i\frac{\theta}{2}} & 0 \\ 0 & e^{i\frac{\theta}{2}} \end{matrix}\right).
\end{equation}
Additionally, we realize a special single-qubit gate $U_s$ with three equal influence samplers:
\begin{equation}
    U_s = \frac{1}{\sqrt3}(\sigma_1+\sigma_2+\sigma_3) =  \frac{1}{\sqrt3}\left( \begin{matrix} 1 & 1-i\\ 1+i & -1 \end{matrix}\right).
\end{equation}
Its influence samplers can be shown as $\mathbb{E}X_l = 1/3$ with $l=1,2,3$. 
\par
As shown in left table in figure~\ref{SM_fig5_Channel}, $X$, $Z$ and $H$ are implemented via a single HWP, while $Y$ and $R_y$ gates require two HWPs. A combination of QWP-HWP-QWP with different angle settings are used to perform $R_x$, $R_z$ and $U_s$ operations. We also implement a non-unitary single-qubit phase damping process, $\Phi_P$, using two phase plates (PPs) with a 3nm filter to control the coherent length of input photons. $\Phi_P$ is described by Kraus operators:
\begin{equation}
    K_1 = \left( \begin{matrix} 1 & 0 \\ 0 & e^{i\phi}\sqrt{1-\lambda}\end{matrix}\right),\ K_2 = \left( \begin{matrix} 0 & 0 \\ 0 & \sqrt{\lambda}\end{matrix}\right),
\end{equation}
where $\lambda$ is the damping rate and $\phi$ is an relative phase between two polarization modes which is determined by the phase plates. 
\par
For two-qubit processes, with control qubit $\ket{q_c}$ and target qubit $\ket{q_t}$, we implement Controlled-NOT ($\mathrm{CNOT}$), Controlled-Z ($\mathrm{CZ}$) and Controlled-$\mathrm{U_s}$ ($\mathrm{CU_s}$), acting on $\ket{q_c}\ket{q_t}$. They have the following forms:
\begin{equation}
    \mathrm{CNOT} = \left( \begin{matrix} 1 & 0 & 0 & 0 \\ 0 & 1 & 0 & 0 \\ 0 & 0 & 0 & 1 \\ 0 & 0 & 1 & 0 \end{matrix}\right),\ \mathrm{CZ} = \left( \begin{matrix} 1 & 0 & 0 & 0 \\ 0 & 1 & 0 & 0 \\ 0 & 0 & 1 & 0 \\ 0 & 0 & 0 & -1 \end{matrix}\right),\ \mathrm{CU_s} = \left( \begin{matrix} 1 & 0 & 0 & 0 \\ 0 & 1 & 0 & 0 \\ 0 & 0 & \frac{1}{\sqrt3} & \frac{1-i}{\sqrt3} \\ 0 & 0 & \frac{1+i}{\sqrt3} & -\frac{1}{\sqrt3} \end{matrix}\right).
\end{equation}
We also implement a two-qubit non-unitary process \textit{Controlled-Phase damping}, denoted as $\Phi_{\mathrm{CP}}$, which is described by the following three Kraus operators
\begin{equation}
    K_1 = \left( \begin{matrix} 1 & 0 & 0 & 0 \\ 0 & 1 & 0 & 0 \\ 0 & 0 & 0 & 0\\ 0 & 0 & 0 & 0\end{matrix}\right),\ K_2 = \left( \begin{matrix} 0 & 0 & 0 & 0 \\ 0 & 0 & 0 & 0 \\ 0 & 0 & 1 & 0\\ 0 & 0 & 0 & e^{i\phi}\sqrt{1-\lambda} \end{matrix}\right),\ K_3 = \left( \begin{matrix} 0 & 0 & 0 & 0 \\ 0 & 0 & 0 & 0 \\ 0 & 0 & 0 & 0\\ 0 & 0 & 0 & \sqrt{\lambda} \end{matrix}\right).
\end{equation}
Figure~\ref{SM_fig5_Channel} details the optical realizations of non-trivial single-qubit ($\Phi_{\{i\}}$) and two-qubit ($\Phi_{\{i,j\}}$) quantum subprocesses. For implementations of two-qubit subprocesses, the path qubit serves as the control and the polarization qubit as the target. Figure~\ref{fig2_IF_result_1} shows the locations of different subprocesses within the experimental setup. $\Phi_{\{2\}}$ ($\Phi_{\{4\}}$) and $\Phi_{\{1,2\}}$ ($\Phi_{\{3,4\}}$) share the same location in the experimental setup, but $\Phi_{\{2\}}$ ($\Phi_{\{4\}}$) is realized by identical optical elements across both path modes, while $\Phi_{\{1,2\}}$ ($\Phi_{\{3,4\}}$) is realized by different elements on each path mode to implement entangled two-qubit processes. Using these subprocesses, along with identity operations, we construct the four-qubit quantum junta processes discussed in this paper.
\section{Inverse test gate implementation and computational basis measurement}
\label{app_i}
\begin{figure}[ht]
    \centering
    \includegraphics[width=0.9\textwidth]{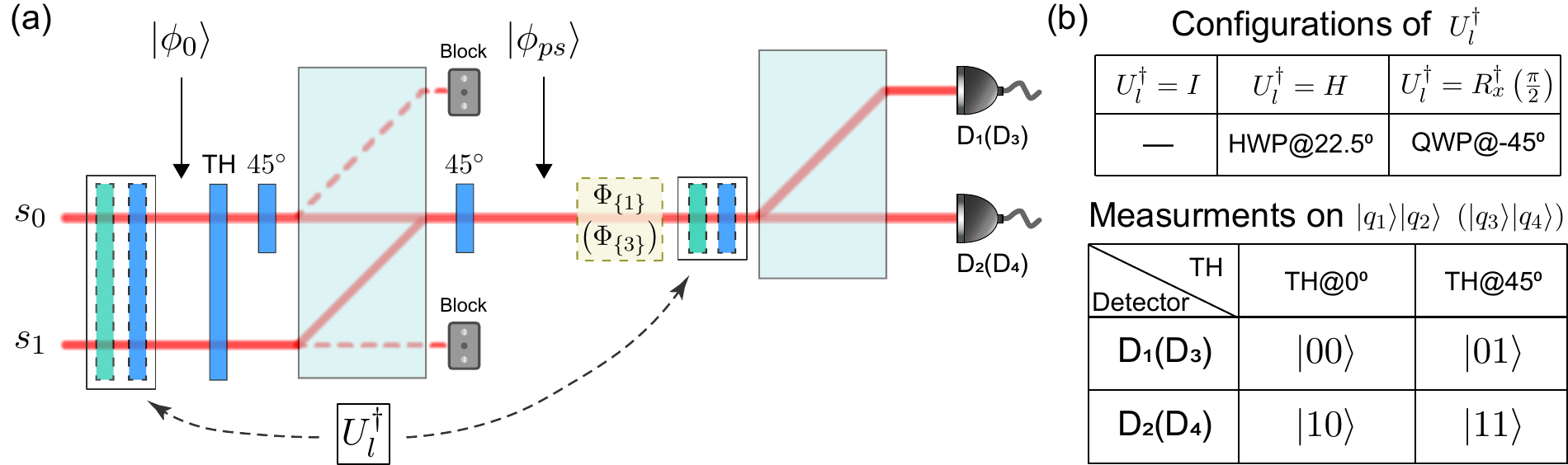}
    \caption{(a) Experimental setup for the inverse test gate $U_{l}^{\dagger}$ and computational basis measurements. TH: toggling HWP. (b) The wave-plate configurations for realizing $U_{l}^{\dagger}$, and the combinations of the toggling HWP and detectors to realize computational basis measurements.}
    \label{SM_fig6_Measure}
\end{figure}
For each photon encoded by a two-qubit state, an inverse test gate $U_l^{\dagger}$ is applied to the polarization qubit after passing a subprocess acting across two path modes, using specific wave-plate configurations, as illustrated in figure~\ref{SM_fig6_Measure}. Since mixed states can be represented as probabilistic mixtures of pure states, we consider a general two-qubit pure state after the first inverse test gate:
\begin{equation}
    \ket{\phi_0} = a\ket{00}+b\ket{01}+c\ket{10}+d\ket{11}.
\end{equation}
Here, the computational basis $\{\ket{00},\ket{01},\ket{10},\ket{11}\}$ is represented by the path and polarization qubits bases $\{\ket{s_0}\ket{H},\ket{s_0}\ket{V},\ket{s_1}\ket{H},\ket{s_1}\ket{V}\}$. 
\par
Computational basis measurements for each photon are performed using a toggling HWP, and a beam splitter (BD) followed by two SPCMs. The test gate $U_l^{\dagger}$ and any subprocess $\Phi_{\{1\}} (\Phi_{\{3\}})$, acting solely on the path qubit (qubits 1 or 3), are applied after the toggling part (including TH, two $45^{\circ}$ HWPs and a BD). We consider a general single-qubit operator $K$ on the path qubit, which can be either a unitary or a Kraus operator.  This operator $K$ represents the combined action of $\Phi_{{1}} (\Phi_{{3}})$ and $U_l^\dagger$ in figure~\ref{SM_fig6_Measure} and takes the form: $K = \left( \begin{matrix} 
    \kappa_{11} & \kappa_{12} \\ 
    \kappa_{21} & \kappa_{22} 
    \end{matrix}\right).$
This operator evolves the input state as follows
\begin{equation}
\begin{split}
    \ket{\phi_1} & = (K\otimes I) \ket{\phi_0} \\
    & = (\kappa_{11}a+\kappa_{12}c)\ket{00}+(\kappa_{11}b+\kappa_{12}d)\ket{01}+(\kappa_{21}a+\kappa_{22}c)\ket{10}+(\kappa_{21}b+\kappa_{22}d)\ket{11}.
\end{split}
\end{equation}
The TH and post-selection, followed by the operation $K$, effectively implement the single-qubit operation $K$ on the path qubit and enable projective measurements in the computational basis. With the TH at $0^{\circ}$, the two-qubit state is projected onto $\ket{00}$ and $\ket{10}$, which are detected by $D_1(D_3)$ for $H$ polarization and $D_2(D_4)$ for $V$ polarization. The evolution of the state $\ket{\phi_0}$ can be expressed as following:
\begin{equation}
\begin{split}
      &  \ket{\phi_0} = \left( \begin{matrix} 
    a \\ 
    b \\
    c \\
    d \\
    \end{matrix}\right) \xrightarrow[]{\text{TH}@0^{\circ}}
    \left( \begin{matrix} 
    a \\ 
    -b \\
    c \\
    -d \\
    \end{matrix}\right) \xrightarrow[\text{placed on } s_0]{\text{HWP}@45^{\circ}}
    \left( \begin{matrix} 
    -b \\ 
    a \\
    c \\
    -d \\
    \end{matrix}\right)\xrightarrow[\text{post-selection}]{\text{BD}}
    \ket{\phi_{ps}}=\left( \begin{matrix} 
    c \\
    a \\
    \end{matrix}\right) \xrightarrow[]{\text{HWP}@45^{\circ}}
    \left( \begin{matrix} 
    a \\
    c \\
    \end{matrix}\right) \\
    & \xrightarrow[]{K}
    \left( \begin{matrix} 
    \kappa_{11}a+\kappa_{12}c \\
    \kappa_{21}a+\kappa_{22}c \\
    \end{matrix}\right) \xrightarrow[\text{Measurement}]{\text{BD \& Detectors}} 
    \begin{cases} 
    D_1(D_3): \text{projection on }\ket{00}\text{ with probability of } |\kappa_{11}a+\kappa_{12}c|^2 \\ 
    D_2(D_4): \text{projection on }\ket{10}\text{ with probability of } |\kappa_{21}a+\kappa_{22}c|^2  
    \end{cases},
\end{split}
\end{equation}
where we use the column vectors denote the amplitudes of quantum states and $\ket{\phi_{ps}}$ results from post-selection and is unnormalized. The TH is set at $45^{\circ}$ to perform the projections onto $\ket{01}$ and $\ket{11}$, as following
\begin{equation}
\begin{split}
    &   \ket{\phi_0} = \left( \begin{matrix} 
    a \\ 
    b \\
    c \\
    d \\
    \end{matrix}\right) \xrightarrow[]{\text{TH}@45^{\circ}}
    \left( \begin{matrix} 
    b \\ 
    a \\
    d \\
    c \\
    \end{matrix}\right) \xrightarrow[\text{placed on } s_0]{\text{HWP}@45^{\circ}}
    \left( \begin{matrix} 
    a \\ 
    b \\
    d \\
    c \\
    \end{matrix}\right)\xrightarrow[\text{post-selection}]{\text{BD}}
    \ket{\phi_{ps}}=\left( \begin{matrix} 
    d \\
    b \\
    \end{matrix}\right) \xrightarrow[]{\text{HWP}@45^{\circ}}
    \left( \begin{matrix} 
    b \\
    d \\
    \end{matrix}\right) \\
    &\xrightarrow[]{K}
    \left( \begin{matrix} 
    \kappa_{11}b+\kappa_{12}d \\
    \kappa_{21}b+\kappa_{22}d \\
    \end{matrix}\right) \xrightarrow[\text{Measurement}]{\text{BD \& Detectors}} 
    \begin{cases} 
    D_1(D_3): \text{projection on }\ket{01}\text{ with probability of } |\kappa_{11}b+\kappa_{12}d|^2 \\ 
    D_2(D_4): \text{projection on }\ket{11}\text{ with probability of } |\kappa_{21}b+\kappa_{22}d|^2  
    \end{cases}.
\end{split}
\end{equation}
The TH and post-selection yield measurement probabilities equivalent to directly measuring $\ket{\phi_1}$ in the computational basis. 
This holds for mixed input states and non-unitary path qubit operations due to the linearity of probabilistic mixtures.
\section{Error analysis and decision threshold}
\label{app_j}
\begin{figure*}[ht]
    \centering
    \includegraphics[width=0.8\textwidth]{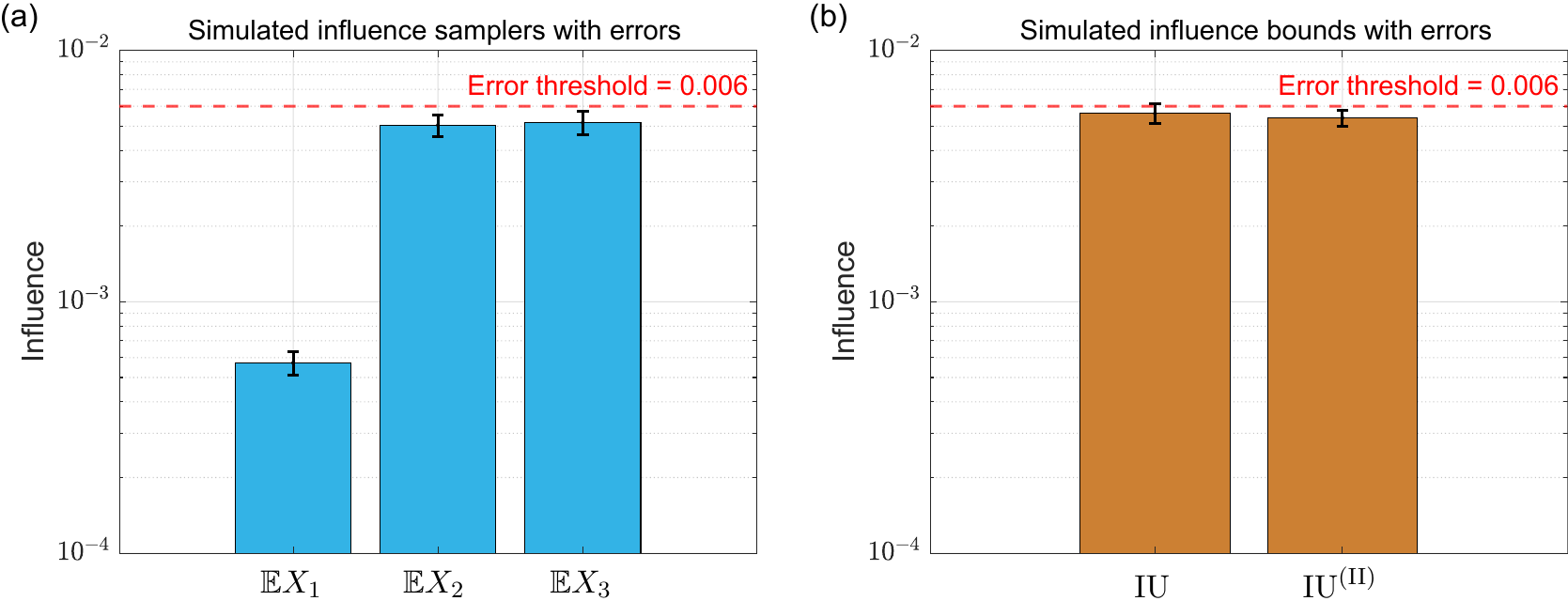}
    \caption{Monte Carlo simulation of (a) influence samplers and (b) influence bounds of a perfect identity process on a path qubit. 100 simulated experimental setups are generated, each with random wave plate errors. For each setup, 100 runs are performed with random errors for wave plate setting angles and random measurement errors. Error bars represent the standard deviation across the 100 setups. The error threshold, $\delta = 0.006$, is consistent with the value used in the actual experiments.
    }
    \label{SM_fig7_IS_simu}
\end{figure*}
In our our experimental setup, state preparation and measurement (SPAM) and test gate errors arise from systematic imperfections in the optical elements. Specifically, the HWP and QWP suffer misalignment of the optics axis (typically $\sim 0.1$ degree), retardation errors (typically $\sim \lambda/300$ where $\lambda=830$nm) and inaccuracies in setting angles (typically $\sim 0.2$ degree). The BDs have an extinction ratio exceeding $1/2000$, and the average interference fringe visibility between the two BDs is approximately 99.8\%. However, the relative phase between the two arms of the BD interferometer drifts during experimental influence sampling, due to vibrations and non-uniformities in the surface of the wave plates across the two path modes. This phase drift significantly affects the path qubits (qubit 1 and qubit 3 in the four-qubit photonic circuit) which are sensitive to the interference. Considering the extinction ratio of BD, imperfect interference visibility, and interference drift, we model the noisy computational basis measurement for path qubits as:
\begin{equation}
    \widetilde{\Pi}_0 = (1-p_e)\ket{0}\!\!\bra{0} + p_e\ket{1}\!\!\bra{1},\ \widetilde{\Pi}_1 = p_e\ket{0}\!\!\bra{0} + (1-p_e)\ket{1}\!\!\bra{1},
\end{equation}
where $p_e\sim 0.0005$ for measuring $\mathbb{E}X_1$ (no interference) and $p_e\sim 0.005$ for measuring $\mathbb{E}X_2$ and $\mathbb{E}X_3$ (with interference). 
\par
Identifying the high-influence qubits utilizes a decision threshold $\delta_0$. To determine a suitable value for $\delta_0$, we perform Monte Carlo simulations of the influence samplers and influence bounds for a perfect identity process (i.e., no optical elements) on a path qubit. This isolates the influence arising solely from systematic errors. We generate 100 simulated experimental setups, each with random retardation and optical axis misalignment errors for the wave plates, sampled from normal distributions with standard deviations equal to the typical error values. For each setup, we run 100 simulations of the influence samplers and bounds, incorporating normally distributed random setting angle errors for the wave plates and exponentially distributed random measurement errors ($p_e$) caused by interference drift. The results are presented in figure~\ref{SM_fig7_IS_simu}.
\par
Our simulations suggest that $\delta_0=0.006$ effectively distinguishes genuine high-influence qubits affected by the quantum process from those primarily affected by SPAM and gate errors. This value, consistent with the threshold used in our experiments, provides a balance between sensitivity to genuine influence and robustness against false identification due to experimental noise.
\begin{figure*}[ht]
    \centering
    \includegraphics[width=1\textwidth]{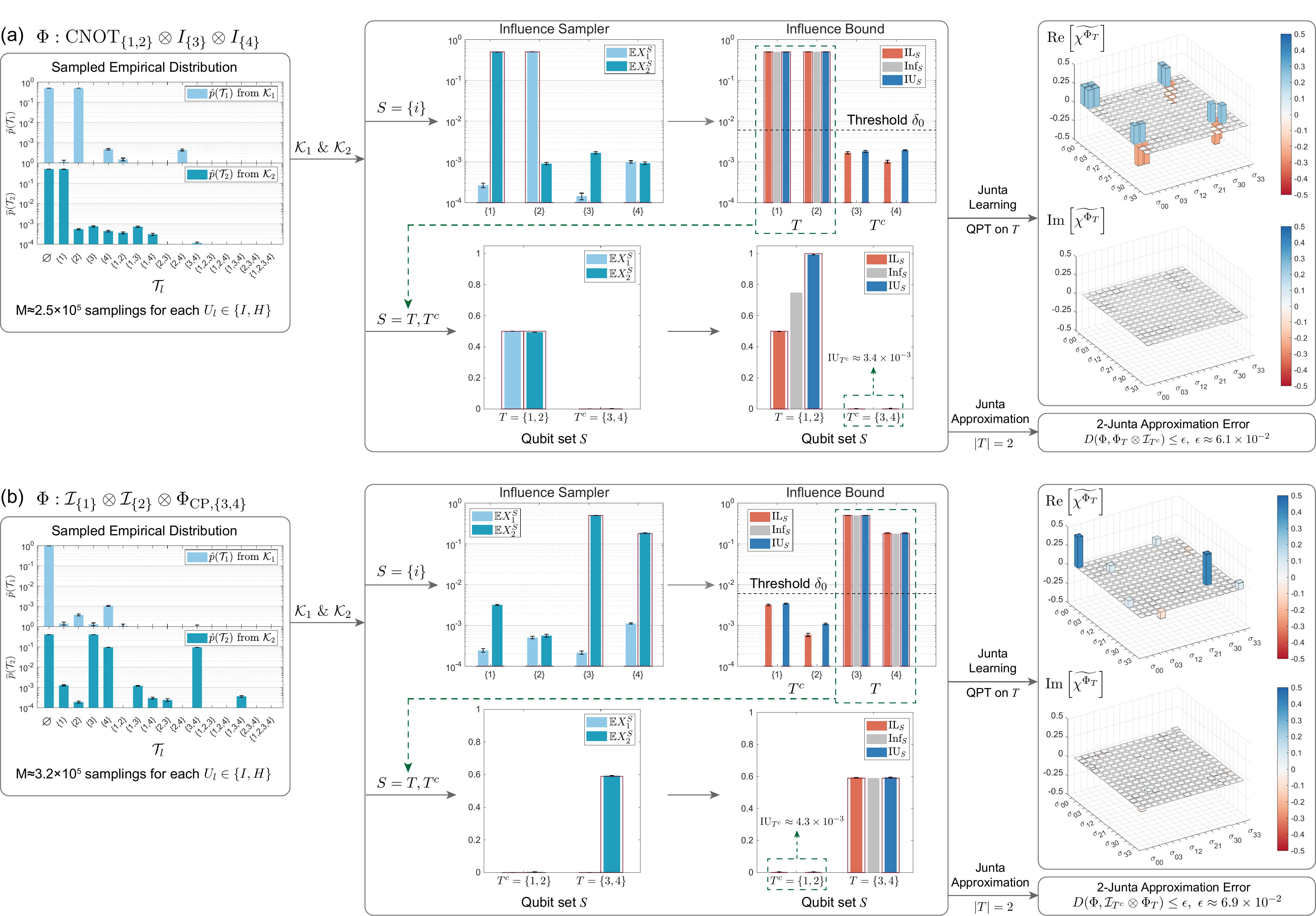}
    \caption{Experimental influence sampling and junta learning for (a) $\mathrm{CNOT}\otimes I\otimes I$ (b) $\mathcal{I}\otimes\mathcal{I}\otimes\Phi_{\mathrm{CP}}$. Here, $\Phi_{\mathrm{CP},\{3,4\}}$ denotes the controlled phase-damping process with $q_3$ as the control qubit and $q_4$ as the target qubit. 
    The decision threshold is $\delta_0=0.006$. 
    The symbols $I$ and $\mathcal{I}$ denote the identity operator and the identity process, respectively.
    }
    \label{SM_fig8_IS_CZ_CP}
\end{figure*}
\section{Supplementary Results}
\subsection*{K.1 Influence sampling on various four-qubit quantum processes}
\label{app_k1}
\begin{figure*}[ht]
    \centering
    \includegraphics[width=1\textwidth]{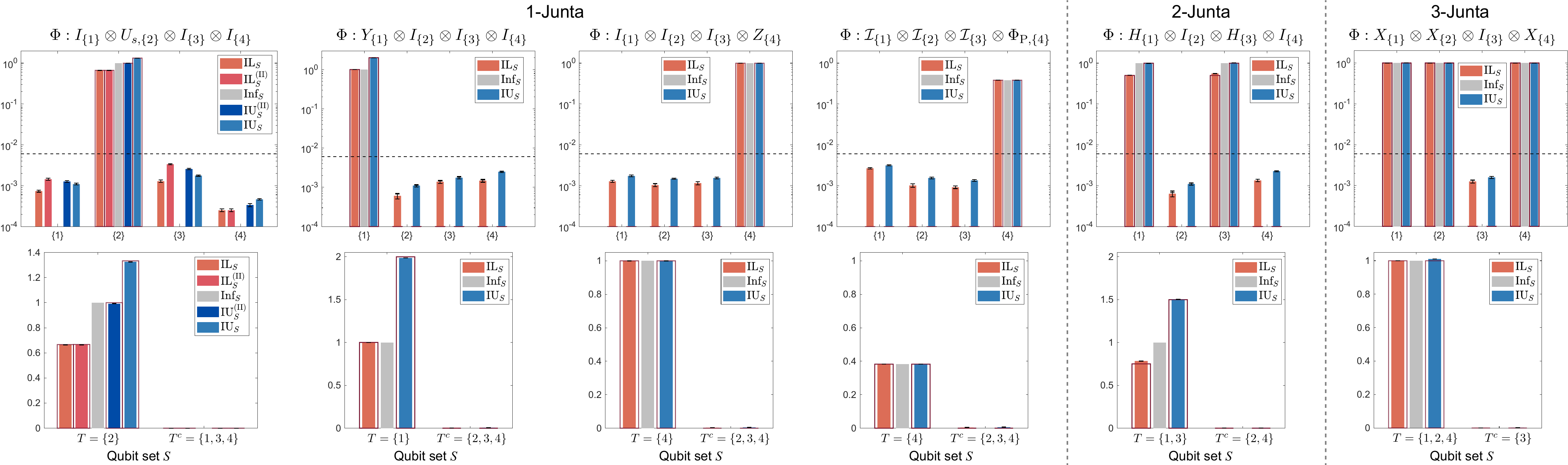}
    \caption{Experimental influence sampling and identification of the high-influence subset $T$ for various quantum junta processes. The black dashed lines indicate the decision threshold $\delta_0=0.006$. $U_s$ denote the special single-qubit gate, and $\Phi_P$ is the phase-damping process with $\phi\approx0.028\pi$ and $\lambda\approx 0.94$. $H$ denotes the Hadamard gate.
    }
    \label{SM_fig9_IS_various}
\end{figure*}
We provide more examples of influence sampling on four-qubit quantum processes, with two test gates $\{I,H\}$, to identify the high-influence two-qubit subsets and to learn the corresponding quantum junta processes, as results shown in figure~\ref{SM_fig8_IS_CZ_CP}.
Two exemplary processes are $\mathrm{CNOT}\otimes I\otimes I$ and $\mathcal{I}\otimes\mathcal{I}\otimes\Phi_{\mathrm{CP}}$, respectively.
\par
The empirical distribution $\hat{p}(\mathcal{T}_l)$ is obtained from $\mathcal{K}_l=\{\mathcal{T}_l^{(j)}\}_{j=1}^{M}$ through $\hat{p}(S)= \mathrm{freq}(S\ \mathrm{in}\ \mathcal{K}_l)/M $, where $S\subseteq [n]$, and $\mathrm{freq}(S\ \mathrm{in}\ \mathcal{K}_l)$ denotes the number of occurrences of $S$ in $\mathcal{K}_l$.
To learn the junta process, we preform the two-qubit quantum process tomography (QPT) on the quantum subprocess on $T$, where $|T|=2$.
In our experiments, the QPT prepares each qubit in $T$ in one of the six eigenstates of the Pauli operators, i.e. $\left\{\ket{0},\ket{1},\ket{+},\ket{-},\ket{L},\ket{R}\right\}^{\otimes 2}$, yielding 36 probe states.
Each qubit is then measured in all six Pauli bases, giving 36 distinct measurement settings.
Therefore, the two-qubit QPT involves $36\times 36$ experimental configurations.
We collect approximately $1.2\times 10^7$ measurement outcomes, from which we obtain a $36\times 36$ empirical probability distribution matrix.
Using convex optimization, we reconstruct a completely positive and trace-preserving process $\widetilde{\Phi_T}$ acting on the $T$, represented by its $\chi$-matrix $\widetilde{\chi^{\Phi_T}}$.
The final learned junta process is then given by $\widetilde{\Phi_T} \otimes \mathcal{I}_{T^c}$.
\par
Influence sampling has also been performed on various other quantum junta processes to identify high-influence qubit subsets, with the results shown in figure~\ref{SM_fig9_IS_various}.
\subsection*{K.2 Influence sampling on a 24-qubit quantum process}
\label{app_k2}
\begin{figure*}[ht]
    \centering
    \includegraphics[width=1\textwidth]{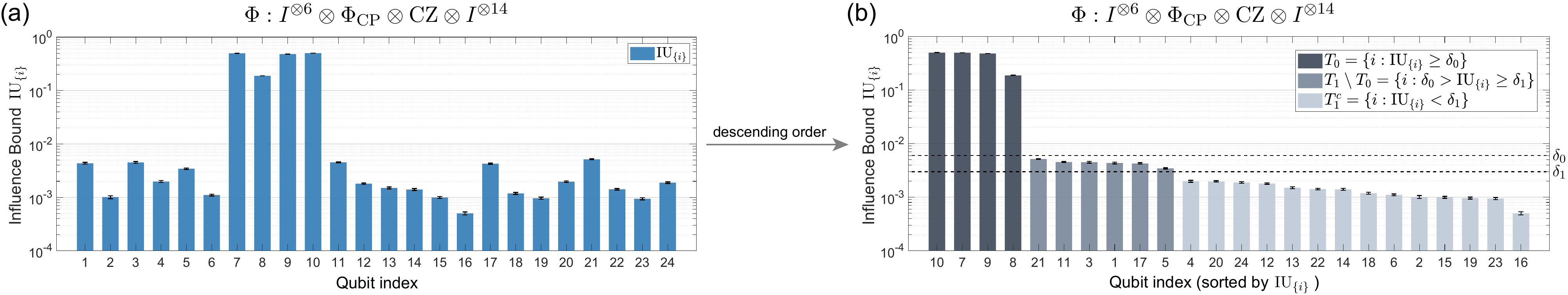}
    \caption{Experimental measured single-qubit influence upper bounds on a 24-qubit quantum process. 
    (a) Qubit ordering by experimental index.
    (b) Qubit ordering by influence bounds in descending order.
    }
    \label{SM_fig10_IS_24}
\end{figure*}
We demonstrate influence sampling on a 24-qubit quantum process by time-multiplexing our four‑qubit device.
Using $N_t$ time modes, in time mode $k_t$ we implement the four-qubit process $\Phi_{[4k_t-3,4k_t]}$ on qubits $4k_t-3$ through $4k_t$, with $k_t\in\{1,2,...,N_t\}$.
Here, we take $N_t=6$, so that the resulting circuit involves 24 qubits in total.
The 24-qubit quantum process is then constructed as $\bigotimes_{k_t=1}^{N_t}\Phi_{[4k_t-3,4k_t]}$, with $\Phi_{[5,8]} = I^{\otimes 2}\otimes \Phi_\mathrm{CP}$, $\Phi_{[9,12]} = \mathrm{CZ}\otimes I^{\otimes 2}$, and $\Phi_{[4k_t-3,4k_t]}=I^{\otimes 4}$ for $k_t\in\{1,4,5,6\}$. Here, $\Phi_\mathrm{CP}$ is the controlled phase-damping process, and $\mathrm{CZ}$ is the controlled-Z gate.
\par
Experimentally, we perform influence sampling on the four-qubit circuit at the time mode $k_t$ to obtain a sample $\T^{(j)}_{l,k_t}$. 
A full sample for the 24-qubit $\Phi$ is then constructed as $\T_l^{(j)}  = \bigcup_{k_t}\T^{(j)}_{l,k_t}$.
Because the chosen subprocesses are independent, the joint distribution $p(\mathcal{T}_l)$ of $\mathcal{T}_l$ factorizes as a product of marginal distributions $p(\mathcal{T}_l) = \prod_{k_t=1}^{N_t}p(\T_{l,k_t})$.
Practically, we firstly estimate the marginal distributions $\hat{p}(\T_{l,k_t})$ using $M>1.3\times10^5$ influence-sampling measurements at each time mode $k_t$.
From these marginals we obtain the single‑qubit influence upper bounds $\mathrm{IU}_{\{i\}}$, with experimental results shown in figure~\hyperref[SM_fig10_IS_24]{10(a)}.
Ordering the qubits by the measured $\mathrm{IU}_{\{i\}}$ in descending order, as shown in figure~\hyperref[SM_fig10_IS_24]{10(b)}, two different decision threshold, $\delta_0=0.006$ and $\delta_1=0.003$, yield two choices of high-influence qubit subsets $T_0$ and $T_1$, with $|T_0|=4$ and $|T_1|=10$.
\par
All marginals $\hat{p}(\T_{l,k_t})$ together form an empirical joint distribution $\hat{p}(\T_{l})=\prod_{k_t=1}^{N_t}\hat{p}(\T_{l,k_t})$.
From the empirical joint distribution $\hat{p}(\T_{l})$ with $l\in\{1,2\}$, we resample datasets $\mathcal{K}_l$, containing $M=10^5$ samples of $\mathcal{T}_l$, using the Monte Carlo method.
The influence upper bounds $\mathrm{IU}_{T_0^c}$ and $\mathrm{IU}_{T_1^c}$ are estimated from $\mathcal{K}_1$ and $\mathcal{K}_2$, yielding $\mathrm{IU}_{T_0^c}=0.0424(8)$ and $\mathrm{IU}_{T_1^c}=0.0178(6)$. 
The corresponding bounds on the junta‑approximation error are $\epsilon_0 = 0.236(3)$ and $\epsilon_1 = 0.146(3)$.
From our results, smaller junta size $|T_0|<|T_1|$ implies larger approximation error $\epsilon_0>\epsilon_1$, revealing the size–error trade‑off of the junta approximation.
\par
In addition, recording only the marginal distribution on each qubit suffices to estimation $\mathrm{IU}_{\{i\}}$, and also yields a looser upper bound on $\Inf[\Phi]_{T^c}$ of the form $\sum_{i\in T^c}\mathrm{IU}_{\{i\}}$, with $\mathrm{IU}_{T^c}\leq \sum_{i\in T^c}\mathrm{IU}_{\{i\}}$.
Our results show that $\sum_{i\in T_0^c}\mathrm{IU}_{\{i\}}=0.0450(4)$ and $\sum_{i\in T_1^c}\mathrm{IU}_{\{i\}}=0.0187(2)$, thereby confirming that $\mathrm{IU}_{T^c}\leq \sum_{i\in T^c}\mathrm{IU}_{\{i\}}$.

\subsection*{K.3 Junta-approximation distance estimation}
\label{app_k3}
\begin{figure*}[ht]
    \centering
    \includegraphics[width=0.5\textwidth]{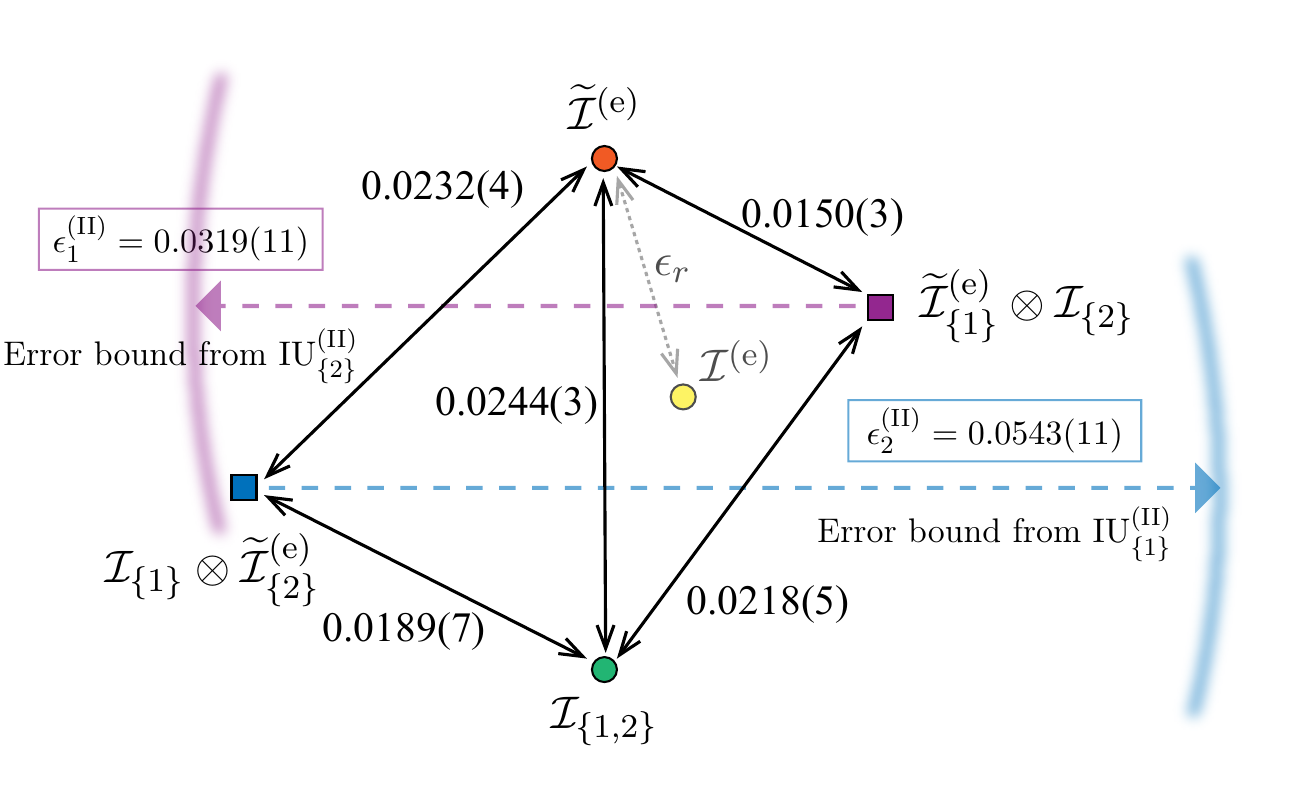}
    \caption{Schematic illustration of the distance relations between the reconstructed process $\widetilde{\mathcal{I}}^{(\mathrm{e})}$, the reconstructed junta approximations $\widetilde{\mathcal{I}}^{(\mathrm{e})}_{\{1\}}\otimes\mathcal{I}_{\{2\}}$ and $\mathcal{I}_{\{1\}}\otimes\widetilde{\mathcal{I}}^{(\mathrm{e})}_{\{2\}}$, and the ideal identity $\mathcal{I}_{\{1,2\}}$. 
    }
    \label{SM_fig11_Dist}
\end{figure*}
To validate the junta-approximation error bound in theorem~\ref{theo_main}, we perform two-qubit QPT on an imperfect two-qubit identity process $\mathcal{I}^{(\mathrm{e})}$. 
The deviation from the ideal identity process arises from experimental imperfections, such as a small phase shift introduced by the beam displacer. 
Although the measurement results are affected by SPAM errors, which may vary across different SPAM configurations, we model the underlying imperfect process $\mathcal{I}^{(\mathrm{e})}$ as fixed, i.e., independent of the choice of probe states and measurements.
For the two-qubit QPT, the probe states are taken from $\left\{\ket{0},\ket{1},\ket{+},\ket{-},\ket{L},\ket{R}\right\}^{\otimes 2}$, and measurements are performed in the two-qubit Pauli bases.
With a total of $1.16\times10^7$ measurement shots, we obtain the outcome statistics as a $36\times 36$ data matrix, from which we reconstruct $\widetilde{\mathcal{I}}^{(\mathrm{e})}$.
\par
Further, we extract the influence sampling datasets from this data matrix, namely those corresponding to state preparation and measurement in eigenstates of the same Pauli operators in $\{\sigma_x^{\otimes2},\sigma_y^{\otimes2},\sigma_z^{\otimes2}\}$.
These datasets are used to estimate the influence bounds $\mathrm{IU}_{T^c}$ and $\mathrm{IU}^{(\mathrm{II})}_{T^c}$ for each choice of $T^c$, namely, $\{1\}$, $\{2\}$ and $\{1,2\}$. 
For a given pair $T$ and $T^c$, we then extract the outcome distributions associated with learning the subprocess $\mathcal{I}^{(\mathrm{e})}_T$ on $T$, i.e., the qubits in $T^c$ are prepared in the maximally mixed state and subsequently traced out. 
Operationally, for $|T|=1$, this is implemented by averaging (mixing) the outcome distributions over the qubit in $T^c$, prepared in the six Pauli bases and measured in the computational basis.
The resulting $6\times 6$ outcome distribution is then used to reconstruct $\mathcal{I}^{(\mathrm{e})}_T$ as $\widetilde{\mathcal{I}}^{(\mathrm{e})}_T$. 
The junta-approximation error is evaluated as $D(\widetilde{\mathcal{I}}^{(\mathrm{e})}, \widetilde{\mathcal{I}}^{(\mathrm{e})}_T\otimes\mathcal{I}_{T^c})$, which can be calculated from the reconstructed $\chi-$matrices.
The error bounds $\epsilon$ and $\epsilon^{(\mathrm{II})}$ are obtained from the measured $\mathrm{IU}_{T^c}$ and $\mathrm{IU}^{(\mathrm{II})}_{T^c}$, respectively. 
For $T=\varnothing$, the junta-approximation is directly given by $\mathcal{I}_{\{1,2\}}$.
The experimental characterized distance relations among the reconstructed processes and the ideal identity are shown in figure~\ref{SM_fig11_Dist}, together with the error bounds from $\mathrm{IU}^{(\mathrm{II})}$. 
The measured influence bounds from influence sampling, the estimated junta-approximation distances, and the corresponding distance bounds are shown in figure~\hyperref[fig3_DB_result_1]{3(a)} and figure~\hyperref[fig3_DB_result_1]{3(b)}.
The results show that the estimated distance lies well within these bounds.
\par
Even though the true process $\mathcal{I}^{\mathrm{(e)}}$ is difficult to determine precisely due to SPAM errors in QPT, i.e., $\epsilon_r$ is unknown, the influence bounds have still provide valid bounds on both $D(\widetilde{\mathcal{I}}^{(\mathrm{e})}, \widetilde{\mathcal{I}}^{(\mathrm{e})}_T\otimes\mathcal{I}_{T^c})$ and $D(\mathcal{I}_{\{1,2\}}, \widetilde{\mathcal{I}}^{(\mathrm{e})}_T\otimes\mathcal{I}_{T^c})$. 
In fact, we expect that $D(\mathcal{I}^{\mathrm{(e)}},\mathcal{I}_{\{1,2\}})<D(\widetilde{\mathcal{I}}^{(\mathrm{e})},\mathcal{I}_{\{1,2\}})$, 
because SPAM errors cause the experimentally reconstructed process $\mathcal{I}^{\mathrm{(e)}}$ to deviate more from the ideal identity than the underlying physical process $\mathcal{I}^{\mathrm{(e)}}$ itself.
\par
To further valid theorem~\ref{theo_main}, we also evaluate the junta-approximation distance and its bound for $\Phi_{\mathrm{CP}}$ using the high-influence subset $T=\{1\}$.
In this case, $\mathrm{IU}_{T^c}^{(\mathrm{II)}}$ is measured as $\mathrm{IU}_{\{2\}}^{(\mathrm{II)}}=0.1735(7)$, producing an error bound $\epsilon^{(\mathrm{II})} \approx 0.54 $. 
The junta-approximation error, estimated via QPT, is $D(\widetilde{\Phi}_{\mathrm{CP}}, \widetilde{\Phi}_{\mathrm{CP},\{1\}}\otimes\mathcal{I}_{\{2\}})\approx 0.18<\epsilon^{(\mathrm{II})}$, which is consistent with the distance bound.


\end{document}